\documentclass[aps,prd,preprint,superscriptaddress,nobibnotes,nofootinbib,a4paper]{revtex4-1}
\usepackage[english]{babel}
\usepackage{graphicx}
\usepackage{amsmath,amssymb}
\usepackage[caption=false]{subfig}
\usepackage{setspace}
\usepackage{hyperref}
\usepackage{rotating}

\newcommand{\p}{\partial}

\newcommand{\Tr}{\mathop{\rm Tr}\nolimits}

\newcommand{\beq}{\begin{eqnarray}}
\newcommand{\eeq}{\end{eqnarray}}
\newcommand{\non}{\nonumber\\}

\begin{document}

\title{Generalized Skyrme model with the loosely bound potential}

\author{Sven Bjarke Gudnason}
\email{bjarke(at)impcas.ac.cn}
\author{Baiyang Zhang}
\email{zhangbaiyang(at)impcas.ac.cn}
\affiliation{Institute of Modern Physics, Chinese Academy of Sciences,
  Lanzhou 730000, China}
\author{Nana Ma}
\email{mann13(at)lzu.edu.cn}
\affiliation{School of Nuclear Science and Technology, Lanzhou
  University, Lanzhou 730000, China}
\affiliation{Institute of Modern Physics, Chinese Academy of Sciences,
  Lanzhou 730000, China}

\date{\today}
\begin{abstract}
We study a generalization of the loosely bound Skyrme model which
consists of the Skyrme model with a sixth-order derivative term --
motivated by its fluid-like properties -- and the second-order loosely
bound potential -- motivated by lowering the classical binding
energies of higher-charged Skyrmions.
We use the rational map approximation for the Skyrmion of topological
charge $B=4$, calculate the binding energy of the latter and estimate
the systematic error in using this approximation. 
In the parameter space that we can explore within the rational map
approximation, we find classical binding energies as low as 1.8\% and
once taking into account the contribution from spin-isospin
quantization we obtain binding energies as low as 5.3\%.
We also calculate the contribution from the sixth-order derivative
term to the electric charge density and axial coupling.

\end{abstract}


\maketitle

\section{Introduction}

The Skyrme model was proposed as a toy model for baryons in a
low-energy effective theory of pions
\cite{Skyrme:1962vh,Skyrme:1961vq}.
The baryon in this theory is identified with the soliton of the
theory -- the Skyrmion.
In the large-$N_c$ limit of QCD this identification is shown by Witten
to be exact \cite{Witten:1983tw,Witten:1983tx}.
Soon after many properties of the nucleon were calculated in the
framework of the (standard) Skyrme model, see
e.g.~Refs.~\cite{Adkins:1983ya,Zahed:1986qz}.
It took a while, however, before progress was made on higher baryon
numbers and the breakthrough came with an approximation using a
rational map \cite{Battye:1997qq,Houghton:1997kg,Battye:2001qn}.
The Skyrmion is a map from point-compactified 3-space, which is
topologically equal to a 3-sphere, to the isospin SU(2), which is also
a 3-sphere.
The rational map approximation\footnote{Throughout the paper, we will
  call it the rational map approximation as opposed to the misleading
  term rational map Ansatz sometimes used in the literature. It is not
  an Ansatz in the sense that the functions of the Ansatz do not
  provide a solution to the field equations. It is an approximation --
  and a rather good one for massless pions -- in that it reproduces
  approximate solutions with only about 1-2\% higher energy than the 
  true solutions. } is an assumption that the 3-sphere can be
factorized into a radial direction ($\mathbb{R}_+$) times a
2-sphere. The latter 2-sphere is then mapped to a 2-sphere in the
target space using the rational map, being a map between Riemann
spheres of degree $B$. The total configuration also has topological
degree $B$, and $B$ is identified with the baryon number. 
This approximation turned out to be a good approximation for a range
of baryons from $B=1$ through $B=22$ -- in the case of massless pions
-- producing fullerene-like structures.

Turning on a physically reasonable pion mass, however, turned out to
induce some alterations
\cite{Battye:2004rw,Battye:2006tb,Houghton:2006ti}; namely, the
fullerenes are no longer the global minimizers of the energy and the
Skyrmions prefer to organize them selves in a crystal made of cubic
4-Skyrmions \cite{Battye:2006na}.
This revelation of the 4-Skyrmion -- which is also the alpha particle
in the model due to unbroken isospin symmetry -- playing an important
role in composing nuclei, turned out to be a welcome feature in the
light of nuclear clustering \cite{Freer:2007}.
The identification of the cluster states in Carbon-12 within the
Skyrme model framework \cite{Lau:2014baa} is one of our main
motivations for using and improving the Skyrme model. 

Many properties of nuclei can be studied after this ground work has
been carried out and new nuclear clusters can be studied.
However, one notorious problem remains; namely the binding energies of
the multi-Skyrmions turn out to be about 1 order of magnitude too
large, compared with experimental data.
This has motivated a line of research trying to modify the Skyrme
model so as to produce much smaller binding energies.
The experimental fact that the binding energies are almost vanishing
led theorists to think that a (deformed) BPS-type model would be a
good candidate for reproducing the low values of the binding
energies. 
The first direction was inspired exactly by this and started off with
a selfdual Yang-Mills theory in 5 dimensions, yielding the Skyrme
model in 4 dimensions with an infinite tower of vector mesons 
\cite{Sutcliffe:2010et,Sutcliffe:2011ig}.
Another proposal came with the discovery of a BPS subsector in the
Skyrme model that is saturable
\cite{Adam:2010fg,Adam:2010ds,Adam:2015ele}, unlike 
the Skyrme-Faddeev bound of the standard Skyrme model
\cite{Skyrme:1962vh,Skyrme:1961vq,Faddeev:1976pg} for which no
solutions saturate it in flat space \cite{Manton:1986pz}\footnote{A
  solution saturating the energy bound exists on a 3-sphere of a
  certain radius \cite{Manton:1986pz}; this is not so useful for
  nuclei though. }. 
This BPS subsector consists of the baryon current squared and a
potential, thus no standard kinetic terms are present in this theory.
However, this sector is integrable and the theory possesses an
infinite amount of symmetries corresponding to volume-preserving
diffeomorphisms. It also has the advantage of modeling a perfect
fluid, which is a welcomed feature in nuclear matter and neutron stars
\cite{Adam:2014nba,Adam:2014dqa,Adam:2015lpa,Adam:2015lra}. 
It is by now called the BPS-Skyrme model.
The problem of perturbing this model is that its near-BPS regime
yields parametrically large field gradients that obviously are very
hard to tackle in numerical calculations using the finite difference
method \cite{Gillard:2015eia,Speight:2014fqa}.
Our motivation for including this term is its fluid-like properties
and that it allows for a limit where the binding energies are small.

In this paper, we follow a third path, inspired by an energy bound
valid for a certain potential \cite{Harland:2013rxa}, which is
basically the standard Skyrme model with a repulsive potential of
fourth-order in $\sigma=\Tr[U]/2$,
\beq
V = \frac{1}{4}m_4^2(1 - \sigma)^4,
\nonumber
\eeq
where $U$ is the field in the usual
chiral Lagrangian \cite{Gillard:2015eia}.
This model was dubbed the lightly bound Skyrme model.
Soon after a better potential was found in
Ref.~\cite{Gudnason:2016mms}, which is of same type but only second
order in $\sigma$
\beq
V = \frac{1}{2}m_2^2(1 - \sigma)^2.
\nonumber
\eeq
By better we mean that it can produce lower binding
energies for the multi-Skyrmions without breaking the platonic
symmetries of the low baryon numbers; in particular, without breaking
the cubic symmetry of the 4-Skyrmion responsible for clustering in the
Skyrme model \cite{Gudnason:2016mms}.
We call this model: the loosely bound Skyrme model and correspondingly
the potential: the loosely bound potential. 
In Ref.~\cite{Gudnason:2016cdo} we have explored the most general
potential up to second order in $\sigma$ and varied the value of the
pion mass in order to find the optimal point in the minimal loosely
bound Skyrme model. In terms of low binding energies, the model
prefers a large pion mass and a large value of the coefficient of the
loosely bound potential. 

\begin{figure}[!ht]
  \begin{center}
    \includegraphics[width=0.49\linewidth]{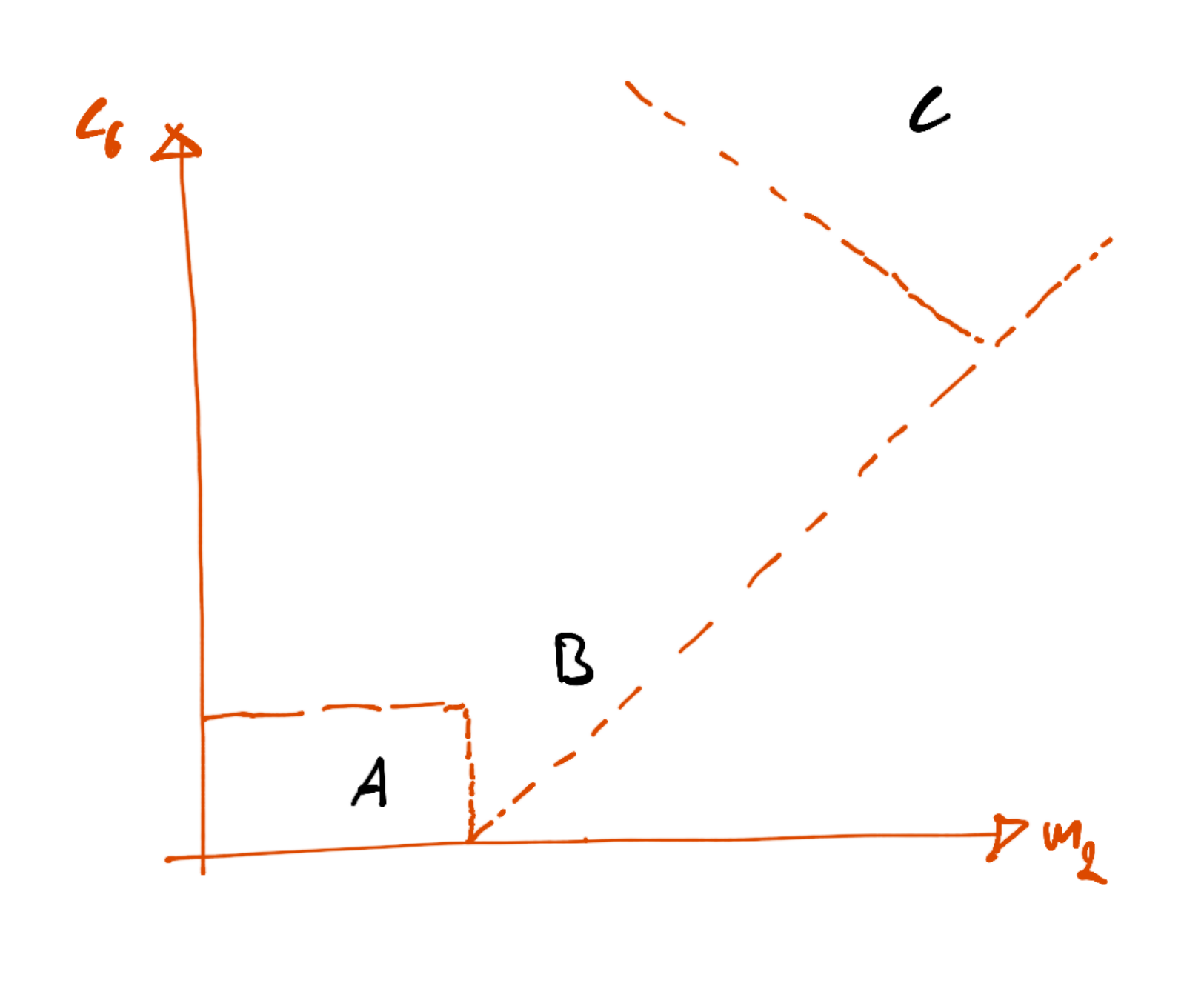}
    \caption{Parameter space in the loosely bound Skyrme model with
      the BPS-Skyrme term. $c_6$ is the BPS-Skyrme term coefficient
      and $m_2$ is the coefficient of the loosely bound potential.
      The region $A$ corresponds to a small BPS-Skyrme term and the
      region of parameter space that we will study in this paper.
      $B$ corresponds to a large $m_2$ and medium-sized value of
      $c_6$.
      Finally, $C$ corresponds to the near-BPS regime of the
      BPS-Skyrme model; in this regime the kinetic term and the Skyrme
      term are small perturbations of the BPS-Skyrme model. }
    \label{fig:parameterspace}
  \end{center}
\end{figure}

In order to improve the remaining issue of too-large binding energies,
we will in this paper include the BPS-Skyrme term. 
The various regimes of the parameter space are sketched in
Fig.~\ref{fig:parameterspace}.

As explained in the figure caption, in region $A$ the BPS-Skyrme term
is relatively small and the normal Skyrme model terms are still
sizable; this is the regime we will study in this paper.
$B$ is the region of parameter space where $m_2$ can be larger due to
the presence of a medium-sized value of $c_6$; this is of course just
our expectation and the exploration of this regime (if it exists),
requires full PDE calculations.
Finally, in regime $C$ both the BPS-Skyrme term and the potential are
huge such that the kinetic term and the normal Skyrme term are mere
perturbations of the BPS-Skyrme model.
This regime is highly nontrivial due to technical problems in the
numerical calculations, as mentioned above.
The region of parameter space to the right of the diagonal dashed line
in the diagram corresponds to very large values of $m_2$ and leads to
Skyrmions that lose the symmetries of platonic solids; in particular 
the 4-Skyrmion loses its cubic symmetry which then becomes
tetrahedral \cite{Gillard:2015eia} and the model in turn loses its
properties of nuclear clustering. 
The tendencies that we explore here in region $A$ will make the
motivations for studying region $B$ in the future; as we will see in
the affirmative.

Let us comment on the relation of the higher-order derivative terms
with respect to the underlying QCD. As QCD at low energies is strongly
coupled, a rigorous explicit derivation is extremely difficult to
carry out.
Nevertheless, the Skyrme term has been derived from the QCD Lagrangian
using partial bosonization where only the phases of the fermions are
bosonized \cite{Zaks:1985cv}; this is done by gauging the flavor
symmetry, however, this gauging does not survive quantization.
A crucial point in this derivation is the claim that the quantum
average of the fermion bilinear in QCD is the same as the quantum
average in the partially bosonized action.
The Skyrme term has been derived in Ref.~\cite{Zaks:1985cv} using this
procedure and in principle higher-order terms can also be considered
this way; in particular the BPS-Skyrme term.
This is, however, beyond the scope of the present paper. 
From a more phenomenologically point of view, the Skyrme term has been
derived from an effective Lagrangian of vector mesons by integrating
out the $\rho$ meson \cite{Ebrahim:1986fb}.
The Skyrme action has also been derived as the low-energy effective
action for the pions in the framework of the Sakai-Sugimoto model
\cite{Sakai:2004cn}; this model is however based on string theory and
is not directly related to QCD.
The BPS-Skyrme term corresponds physically to integrating out the
$\omega$-meson \cite{Adkins:1983nw,Jackson:1985yz} due to the
interaction giving rise to the $\omega\to\pi^+\pi^-\pi^0$ decay. 
To the best of our knowledge, there is no derivation of the loosely
bound (quadratic) potential from QCD as of yet; it is included due to 
its ability to lower the classical binding energies.

The paper is organized as follows.
In Sec.~\ref{sec:model} we present the model and units that we will
be using in this paper. In Sec.~\ref{sec:observables} we calculate all
the observables that we will evaluate.
Sec.~\ref{sec:results} shows the results of the numerical calculations
and evaluation of the observables.
Finally, Sec.~\ref{sec:discussion} concludes with a discussion of the
results found.

\section{The model}\label{sec:model}

The model under consideration is a generalized Skyrme model and the
Lagrangian density in physical units reads
\beq
\mathcal{L} = \frac{f_\pi^2}{4} \mathcal{L}_2
+ \frac{1}{e^2} \mathcal{L}_4
+ \frac{4 c_2 c_6}{c_4^2 f_\pi^2 e^4} \mathcal{L}_6
- \frac{\tilde{m}_\pi^2 f_\pi^2}{4 m_1^2} V,
\eeq
where the kinetic (Dirichlet) term, Skyrme term
\cite{Skyrme:1962vh,Skyrme:1961vq} and BPS-Skyrme 
\cite{Adam:2010fg,Adam:2010ds} term
are given by 
\begin{align}
  \mathcal{L}_2 &= \frac{1}{4}\Tr(L_\mu L^\mu), \qquad
  \mathcal{L}_4 =
  \frac{1}{32}\Tr\left([L_\mu,L_\nu][L^\mu,L^\nu]\right), \non
  \mathcal{L}_6 &=
  \frac{1}{144}\eta_{\mu\mu'}\left(\epsilon^{\mu\nu\rho\sigma}
  \Tr[L_\nu L_\rho L_\sigma]\right)(\epsilon^{\mu'\nu'\rho'\sigma'}
  \Tr[L_{\nu'} L_{\rho'} L_{\sigma'}]),
\end{align}
and the left-invariant current is
\beq
L_\mu \equiv U^\dag \p_\mu U, \qquad
R_\mu \equiv \p_\mu U U^\dag;
\eeq
for later convenience we also defined the right-invariant current
$R_\mu$. 
The constants are the following:
$f_\pi$ is the pion decay constant having units of energy (MeV),
$e>0$ is the dimensionless Skyrme-term coefficient,
$c_6>0$ is the dimensionless BPS-Skyrme term coefficient, 
$\tilde{m}_\pi$ is the pion mass (in MeV) and, finally,
$m_1$ is the dimensionless pion mass parameter.
$c_2$ and $c_4$ are dimensionless constants that can be chosen
arbitrarily; we will fix them shortly. 
$\mu,\nu,\rho,\sigma=0,1,2,3$ are spacetime indices, we are using the
mostly-positive metric signature and $U$ is the Skyrme field which is
related to the pions as 
\beq
U = \mathbf{1}_2\sigma + i\tau^a\pi^a,
\eeq
with $\det U=1$ being the nonlinear sigma model constraint, which is 
equivalent to $\sigma^2+\pi^a\pi^a=1$ and $\tau^a$ are the Pauli
matrices.

We will now rescale the theory and work in (dimensionless) Skyrme
units, following Ref.~\cite{Gudnason:2016cdo}. 
The lengths are rescaled as $\tilde{x}^i=\mu x^i$, where both
$\tilde{x}^i$ and $\mu$ have units of inverse energy (MeV${}^{-1}$),
and similarly the energy is rescaled as $\tilde{E}=\lambda E$; where
$\tilde{E}$ and $\lambda$ have units of energy (MeV).
Finally we get the dimensionless Lagrangian density
\beq
\mathcal{L} = c_2 \mathcal{L}_2 + c_4 \mathcal{L}_4
+ c_6 \mathcal{L}_6 - V,
\label{eq:L}
\eeq
where $c_2>0$ and $c_4>0$ are positive definite real constants and
$c_6\geq 0$ is a positive semi-definite real constant and the
rescaling parameters are determined as
\beq
\lambda = \frac{f_\pi}{2e\sqrt{c_2c_4}}, \qquad
\mu = \sqrt{\frac{c_2}{c_4}} \frac{2}{e f_\pi},
\label{eq:unit_conversion}
\eeq
whereas the pion mass in physical units (MeV) is given by
\beq
\tilde{m}_\pi = \frac{\sqrt{c_4}}{2c_2} e f_\pi m_1.
\label{eq:mpion_physical_units}
\eeq
We have assumed in the above expression, that the part of the
potential $V$ contributing to the pion mass is normalized to $m_1$
in dimensionless units.

A comment about the normalization chosen for the BPS-Skyrme term,
$\mathcal{L}_6$, in this paper is in store.
Derrick's theorem \cite{Derrick:1964ww} implies that a
higher-derivative term is necessary for stability of a soliton with
finite size and energy.
Skyrme used the simplest possibility, namely, a fourth-order
derivative correction to stabilize the Skyrmion, but a sixth-order
term like the BPS-Skyrme term can equally well stabilize the Skyrmion
without the presence of the standard Skyrme term.
This situation corresponds to $c_4=0$ and $c_6>0$, which however is
not possible in our normalization and calibration of the energy and
length units.
As explained in the introduction, in this paper we focus on the region
in parameter space where the BPS-Skyrme term is added as a
perturbation to the loosely bound Skyrme model.

The potential we will use in this paper is composed by the loosely
bound potential \cite{Gudnason:2016mms} and the standard pion mass
term 
\beq
V = V_1 + V_2,
\label{eq:V}
\eeq
where we have defined
\beq
V_n \equiv \frac{1}{n} m_{n}^2 (1 - \sigma)^n.
\eeq
This is the part of the most general potential to second order in
$\sigma=\Tr[U]/2$ giving the lowest binding energies
\cite{Gudnason:2016cdo}.

The Lagrangian density \eqref{eq:L} without a potential, possesses
$\mathrm{SU}(2)\times\mathrm{SU}(2)$ symmetry, which is explicitly
broken to the diagonal SU(2) by the potential \eqref{eq:V}. The latter
SU(2) corresponds to isospin which we will keep unbroken in this
paper. 

The Skyrme field is a map from point-compactified 3-space
$\mathbb{R}^3\cup\{\infty\}\simeq S^3$ to the target space, SU(2),
characterized by the third homotopy group
\beq
\pi_3(\textrm{SU}(2)) = \mathbb{Z} \ni B,
\eeq
where $B$ is the topological degree, also called the baryon number.
The baryon number is the integral of the baryon charge density
\beq
B = \frac{1}{2\pi^2}\int d^3x\; \mathcal{B}^{0},
\eeq
with
\beq
\mathcal{B}^0 = -\frac{1}{12}\epsilon^{ijk} \Tr[L_i L_j L_k].
\label{eq:Bdensity}
\eeq
Throughout the paper we will use the short-hand notation $B$-Skyrmion
for a Skyrmion with topological degree $B$.

For convenience we allow for a generic normalization of the terms in
the model. However, once we want to calibrate the model, we have to
fix the choice of the coefficients $c_2$ and $c_4$. 
The standard choice of Skyrme units corresponds to setting $c_2=c_4=2$
for which energies and lengths are given in units of $f_\pi/(4e)$ and 
$2/(ef_\pi)$, respectively \cite{MantonSutcliffe}. 
In this paper, we will adhere to the convention used in
Refs.~\cite{Gudnason:2016mms,Gudnason:2016cdo}, i.e.,
\beq
c_2 = \frac{1}{4}, \qquad
c_4 = 1,
\label{eq:c2c4}
\eeq
for which, according to Eq.~\eqref{eq:unit_conversion}, energies and
lengths are given in units of $f_\pi/e$ and $1/(ef_\pi)$,
respectively.  
The pion mass in physical units is
\beq
\tilde{m}_\pi = \frac{\sqrt{c_4}}{2c_2} e f_\pi
\left.\sqrt{-\frac{\p V}{\p \sigma}}\right|_{\sigma=1}
= 2 e f_\pi m_1,
\label{eq:physical_mp}
\eeq
where we have used the normalization \eqref{eq:c2c4} in the last
equality. 
The pion mass $m=1$ used in Ref.~\cite{Battye:2006na}, corresponds to 
$m_1=1/4$ and in turn $\tilde{m}_\pi=e f_\pi/2$ in our units and
normalization.

\section{Observables}\label{sec:observables}

We have now defined the model, set the notation and we are ready
to calculate the expressions for the observables that we will
determine numerically and compare to experimental data.
We will follow Ref.~\cite{Gudnason:2016cdo} and calculate Skyrmions of 
baryon numbers one and four; this choice is advantageous for several
reasons: the importance of the 4-Skyrmion ($B=4$) as it corresponds to
the alpha particle and plays a crucial role in nuclear clustering;
the ground state of ${}^4$He is a spin-0, isospin-0 state, which means
that it is determined by the classical energy of the 4-Skyrmion and
finally, the 1-Skyrmion represents the nucleon/proton/neutron in the
theory with unbroken isospin and therefore we can use this for
comparison with experimental data for the proton. 

The 1-Skyrmion is spherically symmetric and hence is described by the
hedgehog Ansatz
\beq
U^{(1)} = \mathbf{1}_2\cos f(r) + i\hat{x}^a\tau^a\sin f(r),
\label{eq:hedgehog}
\eeq
where $r=\sqrt{(x^1)^2+(x^2)^2+(x^3)^2}$ is the radial coordinate and
$\hat{x}^a=x^a/r$ is the spatial unit 3-vector.
The classical energy for the 1-Skyrmion is given by plugging the above
Ansatz into the Lagrangian \eqref{eq:L} yielding
\begin{align}
  E_1 &= -\int d^3x\;\mathcal{L}\left[U^{(1)}\right] \non
  &= 4\pi\int dr\; r^2\bigg[c_2\left(\frac{1}{2} f_r^2
  +\frac{1}{r^2}\sin^2 f\right)
  + c_4\frac{\sin^2f}{r^2}\left(f_r^2 + \frac{\sin^2f}{2r^2}\right)
  + c_6\frac{\sin^4(f) f_r^2}{r^4} \non
  &\phantom{=4\pi\int dr\;r^2\bigg[\ }
  + m_1^2(1-\cos f)
  + \frac{1}{2} m_2^2 (1-\cos f)^2\bigg],
  \label{eq:E1}
\end{align}
where $f_r\equiv\p_r f$.

\subsection{Calibration}

In this paper we choose like in
Refs.~\cite{Gudnason:2016mms,Gudnason:2016cdo} to calibrate the model
by fitting the mass and the size of ${}^4$He to those of the $B=4$
Skyrmion. 
Since the numerical calculations in the $B=4$ sector are numerically
expensive, we choose to use the rational map approximation to estimate 
the mass and size of the 4-Skyrmion. 
It is known that in the standard Skyrme model, the rational map
approximation provides quick solutions within about 1 percent
accuracy \cite{Houghton:1997kg,Battye:2001qn}.
This will, however, be the first time it is used in the 
loosely bound Skyrme model; we will therefore check the results in the
$c_6=0$ sector by comparing to the full PDE solutions of
Ref.~\cite{Gudnason:2016cdo}. 

The rational map is made by performing a radial suspension of the
Skyrme field
\beq
U^{({\rm RM})} = \mathbf{1}\cos f + i\tau^a n^a\sin f,
\eeq
where $n^a$ is a unit 3-vector.
The suspension means choosing $f(r)$ and $n(\theta,\phi)$ in normal
spherical coordinates.
It will however be convenient to use the complex coordinate
$z=e^{i\phi}\tan\frac{\theta}{2}$ on the Riemann sphere, for which the
2-sphere reads
\beq
n^1 = \frac{z + \bar{z}}{1 + |z|^2}, \qquad
n^2 = \frac{i(\bar{z} - z)}{1 + |z|^2}, \qquad
n^3 = \frac{1 - |z|^2}{1 + |z|^2}.
\eeq
The generalization to a degree $B$ map on the Riemann sphere is then
made by using the rational map $R(z)$, which is a holomorphic function
of $z$ and the 3-vector thus reads \cite{Battye:2001qn}
\beq
n^1 = \frac{R(z) + \bar{R}(\bar{z})}{1 + |R(z)|^2}, \qquad
n^2 = \frac{i(\bar{R}(\bar{z}) - R(z))}{1 + |R(z)|^2}, \qquad
n^3 = \frac{1 - |R(z)|^2}{1 + |R(z)|^2}.
\eeq
We can now write our static Lagrangian density as
\cite{Battye:2001qn,Kopeliovich:2001ca,Adam:2010fg,Adam:2010ds}
\begin{align}
  -\mathcal{L}\left[U^{({\rm RM})}\right] &=
  c_2\left(\frac{1}{2} f_r^2
  + \frac{\sin^2 f}{r^2}\frac{(1+|z|^2)^2}{(1+|R|^2)^2} |R_z|^2\right)
  \non
  &\phantom{=\ }
  +c_4\frac{\sin^2 f}{r^2}\left(f_r^2\frac{(1+|z|^2)^2}{(1+|R|^2)^2}
  |R_z|^2
  +\frac{\sin^2 f}{2r^2} \frac{(1+|z|^2)^4}{(1+|R|^2)^4} |R_z|^4\right)
  \non
  &\phantom{=\ }
  +c_6 \frac{\sin^4(f)f_r^2}{r^4}\frac{(1+|z|^2)^4}{(1+|R|^2)^4}
  |R_z|^4
  + m_1^2(1 - \cos f) + \frac{1}{2}m_2^2(1-\cos f)^2,
\end{align}
which by integration over 3-space gives
\begin{align}
  E_B^{({\rm RM})} &= -\int d^3x\; \mathcal{L}\left[U^{({\rm RM})}\right] \non
  &= 4\pi\int dr \bigg[
    c_2\left(\frac{1}{2}r^2 f_r^2 + B\sin^2 f\right) 
    +c_4\sin^2 f\left(B f_r^2
    + \mathcal{I}\frac{\sin^2 f}{2r^2}\right)
    +c_6\mathcal{I}\frac{\sin^4(f)f_r^2}{r^2} \non
&\phantom{=4\pi\int dr\bigg[\ }
    +m_1^2(1-\cos f)
    +\frac{1}{2}m_2^2(1-\cos f)^2\bigg],
  \label{eq:EB}
\end{align}
where
\beq
\mathcal{I}\equiv \frac{1}{4\pi}
\int\frac{2idz\wedge d\bar{z}}{(1+|z|^2)^2}
\left(\frac{1+|z|^2}{1+|R|^2} R_z\right)^4.
\eeq

For the 4-Skyrmion the minimizing rational map has the form
\cite{Houghton:1997kg}
\beq
R(z) = \frac{z^4 + i2\sqrt{3} z^2 + 1}{z^4 - i2\sqrt{3} z^2 + 1},
\label{eq:R4}
\eeq
which upon integration gives
$\mathcal{I}_{4}\simeq 20.6496$.

With the (rational map) approximated solution for the cubic
4-Skyrmion, we can now perform the calibration by calculating the mass
and size of the solution.
Since, as we mentioned already, the ground state of ${}^4$He is a
spin-0, isospin-0 state, the mass does not receive a contribution from
spin-isospin quantization and is thus given by the classical static
energy \eqref{eq:EB}.
The electric charge density is given by
$\rho_E=\frac{1}{2}\frac{1}{2\pi^2}\mathcal{B}^0 + \mathcal{I}^3$
\cite{Adkins:1983ya}, but since the isospin-0 state does not
contribute to the charge density, the charge radius is determined only
from the baryon charge density
\beq
\mathcal{B}^0 = -\frac{B\sin^2(f)f_r}{r^2},
\eeq
which yields the baryon charge and electric charge radii (squared)
\beq
r_4^2 = r_{E,4}^2 = -\frac{2}{\pi}\int dr\; r^2
\sin^2(f)f_r,
\eeq
where $f$ is a minimizer of the energy \eqref{eq:EB} with the rational
map \eqref{eq:R4} and we have normalized the integral by dividing with
$B/2$.

We can finally determine the parameters of the model as
\beq
f_\pi = 2\sqrt{c_2}
\sqrt{\frac{r_{E,4} M_{{}^4{\rm He}}}{r_{{}^4{\rm He}} E_4}}, \qquad
e = \frac{1}{\sqrt{c_4}}
\sqrt{\frac{r_{E,4} E_4}{r_{{}^4{\rm He}} M_{{}^4{\rm He}}}},
\eeq
which simplifies with the convention \eqref{eq:c2c4} to
\beq
f_\pi = \sqrt{\frac{r_{E,4} M_{{}^4{\rm He}}}{r_{{}^4{\rm He}} E_4}}, \qquad
e = \sqrt{\frac{r_{E,4} E_4}{r_{{}^4{\rm He}} M_{{}^4{\rm He}}}},
\label{eq:fpi_e_calib}
\eeq
where it is understood that $r_{E,4}=\sqrt{r_{E,4}^2}$ and the values
of the experimental data used are
$M_{{}^4{\rm He}}=3727$ MeV and
$r_{{}^4{\rm He}}=8.492\times 10^{-3}$ MeV${}^{-1}$.

\subsection{Mass spectrum}

We are now ready to calculate the mass spectrum; the first mass we
need is the mass of the nucleon.
It has two contributions: the (classical) soliton mass, $E_1$
(see Eq.~\eqref{eq:E1}), and that coming from spin-isospin
quantization, $\epsilon_1$ which we will now calculate. 

The quantum contribution from spin-isospin quantization can be
calculated from the kinetic part of the Lagrangian \eqref{eq:L}, as
follows 
\beq
T = \frac{1}{2} a_i U_{ij} a_j
= \Lambda \Tr[\dot{A}\dot{A}^\dag], \qquad
a_i \equiv -i\Tr(\tau^i A^\dag\dot{A}).
\eeq
Because of spherical symmetry of the nucleon, we get
$U_{ij}=\Lambda\delta_{ij}$ with \cite{Adkins:1983ya,Dube:1989gu,Marleau:1990nh}
\begin{align}
  \Lambda = \frac{8\pi}{3}\int dr\; r^2\sin^2 f\left[
    c_2 + c_4 f_r^2 + \frac{c_4}{r^2}\sin^2 f
    +\frac{2c_6\sin^2(f) f_r^2}{r^2}\right].
\end{align}
Since $\Tr[\dot{A}\dot{A}^\dag]$ is the kinetic term on the 3-sphere,
the quantization thereof yields
\beq
T = \frac{1}{8\Lambda}\ell(\ell+2)
= \frac{1}{2\Lambda} J(J+1),
\eeq
where $J=\ell/2$ is the spin quantum number.
Finally, the spin contribution for the spin-1/2 ground state of the
nucleon reads 
\beq
T_{1/2} = \frac{1}{2\Lambda}\frac{3}{4},
\eeq
and in physical units
\beq
\tilde{m}_N = 
\tilde{E}_1 + \tilde{\epsilon}_1 =
\frac{f_\pi}{e} E_1 + \frac{3e^3f_\pi}{8\Lambda}.
\eeq

Now we can quickly get the mass of the Delta resonance, by setting
$J=3/2$ and hence we get
\beq
\tilde{m}_\Delta = 
\tilde{E}_1 + 5\tilde{\epsilon}_1.
\eeq

The final mass that we will calculate and compare to data in this
paper is the pion mass, which is given in physical units in
Eq.~\eqref{eq:physical_mp}.

\subsection{Binding energy}

One of the prime observables in this paper is the binding energy. 
The classical and total binding energies are defined as
\beq
\Delta_B = B E_1 - E_B, \qquad
\Delta_B^{\rm tot} = B(E_1 + \epsilon_1) - E_B - \epsilon_B,
\eeq
and the relative classical and total binding energies in turn read 
\beq
\delta_B = 1 - \frac{E_B}{B E_1}, \qquad
\delta_B^{\rm tot} = 1 - \frac{E_B + \epsilon_B}{B (E_1 + \epsilon_1)},
\eeq
respectively.
Notice that the quantum contribution to the $B$-Skyrmion,
$\epsilon_B$, \emph{lowers} the total binding energy, whereas the
quantum contribution to the 1-Skyrmion \emph{raises} the total binding 
energy.
In particular, for the 4-Skyrmion, the ground state is a spin-0,
isospin-0 state and thus the quantum contribution vanishes,
i.e.~$\epsilon_4=0$, yielding the relative total binding energy
\beq
\delta_4^{\rm tot} = 1 - \frac{E_4}{B (E_1 + \epsilon_1)}.
\eeq
This means that for the 4-Skyrmion, the spin-isospin quantization only
exacerbates the problem of the binding energy being too large.

\subsection{Charge radii}

We will now calculate the baryon charge radius of the nucleon and the
electric charge radius of the proton.
We begin by calculating the vectorial current.
The infinitesimal transformation
\beq
U \to U + i\theta_V^a[Q^a,V],
\eeq
thus gives rise to the vectorial current
\begin{align}
  J_{V}^{\mu a} &=
  \frac{ic_2}{2} \Tr\left[(R^\mu-L^\mu)Q^a\right]
  +\frac{ic_4}{8} \Tr\left[
    \left([R_\nu,[R^\mu,R^\nu]] - [L_\nu,[L^\mu,L^\nu]]\right)
    Q^a\right] \non
  &\phantom{=\ }
  +\frac{ic_6}{24}\eta_{\nu\nu'}\epsilon^{\nu'\rho'\sigma'\tau'}
    \Tr[L_{\rho'} L_{\sigma'} L_{\tau'}]
    \epsilon^{\nu\rho\sigma\mu}\Tr\left[
      \left(R_\rho R_\sigma - L_\rho L_\sigma\right) Q^a\right],
\end{align}
whose integrated zeroth component for the hedgehog Ansatz
\eqref{eq:hedgehog}, can be written as 
\begin{equation}
\int d^3x\; J_{V}^{0a} =
-\frac{i8\pi}{3}
\int dr\left[
  c_2 r^2\sin^2 f
  +c_4\sin^2 f\left(\sin^2 f + r^2 f_r^2\right)
  +2c_6\sin^4(f) f_r^2\right]
\Tr[\tau^a \dot{A}A^\dag], \nonumber
\end{equation}
from which we can construct the electric (radial) density as
\cite{Adkins:1983ya} 
\begin{align}
  \int d\Omega\;\rho_E &= \int d\Omega\left(
  \frac{1}{2}\frac{1}{2\pi^2}\mathcal{B}^0 + \mathcal{I}^3 \right)\non
  &= -\frac{\sin^2(f) f_r}{\pi r^2}
  + \frac{c_2\sin^2 f + \frac{c_4}{r^2}\sin^2(\sin^2f + r^2 f_r^2) +
    \frac{2c_6}{r^2}\sin^4(f) f_r^2}{2\int dr\left(c_2r^2\sin^2 f
    + c_4\sin^2(\sin^2f + r^2 f_r^2) + 2c_6\sin^4(f) f_r^2\right)},
\end{align}
where we have used $Q^a=\tau^a/2$.
The above radial function is the charge density of the proton and it 
integrates to unity, i.e.~$\int dr\; r^2\rho_E=1$ (as it should). 
We can thus construct the electric charge radius (squared) as
\beq
r_{E,1}^2 = \int dr\; r^4 \rho_E,
\label{eq:rE1}
\eeq
whereas the baryon charge radius is given by
\beq
r_1^2 = -\frac{2}{\pi}\int dr\; r^2\sin^2(f) f_r.
\label{eq:rB1}
\eeq

\subsection{Axial coupling}

In this paper we will consider the axial coupling
\cite{Adkins:1983ya}, in addition to the ones considered in
Ref.~\cite{Gudnason:2016cdo}.
The axial current reads
\begin{align}
  J_{A}^{\mu a} &=
  \frac{ic_2}{2} \Tr\left[(R^\mu+L^\mu)Q^a\right]
  +\frac{ic_4}{8} \Tr\left[
    \left([R_\nu,[R^\mu,R^\nu]] + [L_\nu,[L^\mu,L^\nu]]\right)
    Q^a\right] \non
  &\phantom{=\ }
  +\frac{ic_6}{24}\eta_{\nu\nu'}\epsilon^{\nu'\rho'\sigma'\tau'}
    \Tr[L_{\rho'} L_{\sigma'} L_{\tau'}]
    \epsilon^{\nu\rho\sigma\mu}\Tr\left[
      \left(R_\rho R_\sigma + L_\rho L_\sigma\right) Q^a\right],
\end{align}
corresponding to the infinitesimal transformation
\beq
U \to U + i\theta_A^a\{Q^a,U\}.
\eeq
Imposing spherical symmetry and ignoring second-order in
time-derivatives, we can write 
\beq
\int d^3x \; J_{Ai}^a = \frac{g_A}{2} \Tr\left[\tau^i A^\dag \tau^a A\right],
\eeq
where the axial coupling is given by\footnote{Although the contributions to
  the axial coupling from the terms up to sixth order in derivatives
  have been calculated in Refs.~\cite{Dube:1989gu,Marleau:1990nh} our
  expression does not agree with that of the latter references.
  However, our expression agrees with that of
  Ref.~\cite{Adkins:1983ya} up to fourth-order derivative terms. }
\begin{align}
  g_A &= - \frac{4\pi}{3}\int dr\; r\bigg[
    c_2\left(\sin 2f + r f_r\right)
    +c_4\left(\frac{\sin^2 f\sin 2f}{r^2}
    +\frac{2\sin^2(f) f_r}{r}
    +\sin(2f) f_r^2\right) \non
&\phantom{=-\frac{16\pi}{3}\int dr\;r\bigg[\ }
    +\frac{2c_6\sin^2 f}{r^2}\left(\frac{\sin^2(f)f_r}{r}
    +\sin(2f) f_r^2\right)
    \bigg],
\end{align}
and we have used $Q^a=\tau^a/2$.
Although the axial coupling is dimensionless, it is still in Skyrme
units and so to relate to its experimentally observed value, we need
to change the units back to physical units
\beq
\tilde{g}_A = \frac{g_A}{e^2c_4} = \frac{g_A}{e^2}
\label{eq:gAtilde}
\eeq
by multiplying the result by $\lambda\mu=1/(e^2c_4)$ and in the last
equality we have used the convention of Eq.~\eqref{eq:c2c4} which sets
$c_4=1$.

\section{Results}\label{sec:results}

We have now presented the loosely bound Skyrme model with up to six
orders of derivative terms and the observables that we will compare to
experimental data.
Now we just need to calculate the numerical solutions of the Skyrmion
profile functions for the 1-Skyrmion and the 4-Skyrmion, respectively,
in order to evaluate the observables presented in
Sec.~\ref{sec:observables}.
We calculate the radial ODEs for the Skyrmion profile functions using
the relaxation method and show the results below.

\subsection{Binding energies}

\begin{figure}[!htp]
  \begin{center}
    \mbox{
      \subfloat[]{\includegraphics[width=0.49\linewidth]{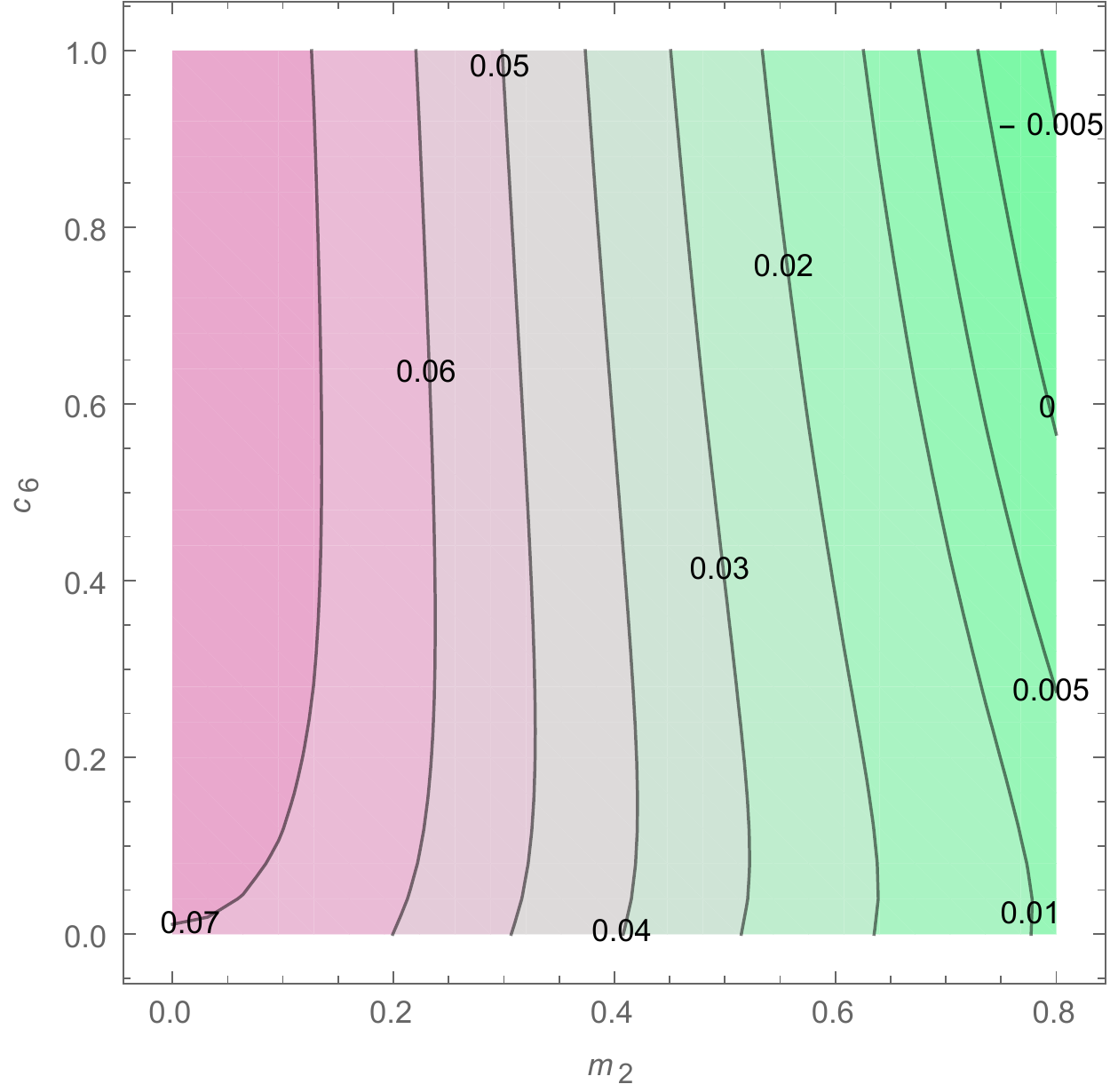}}
      \subfloat[]{\includegraphics[width=0.49\linewidth]{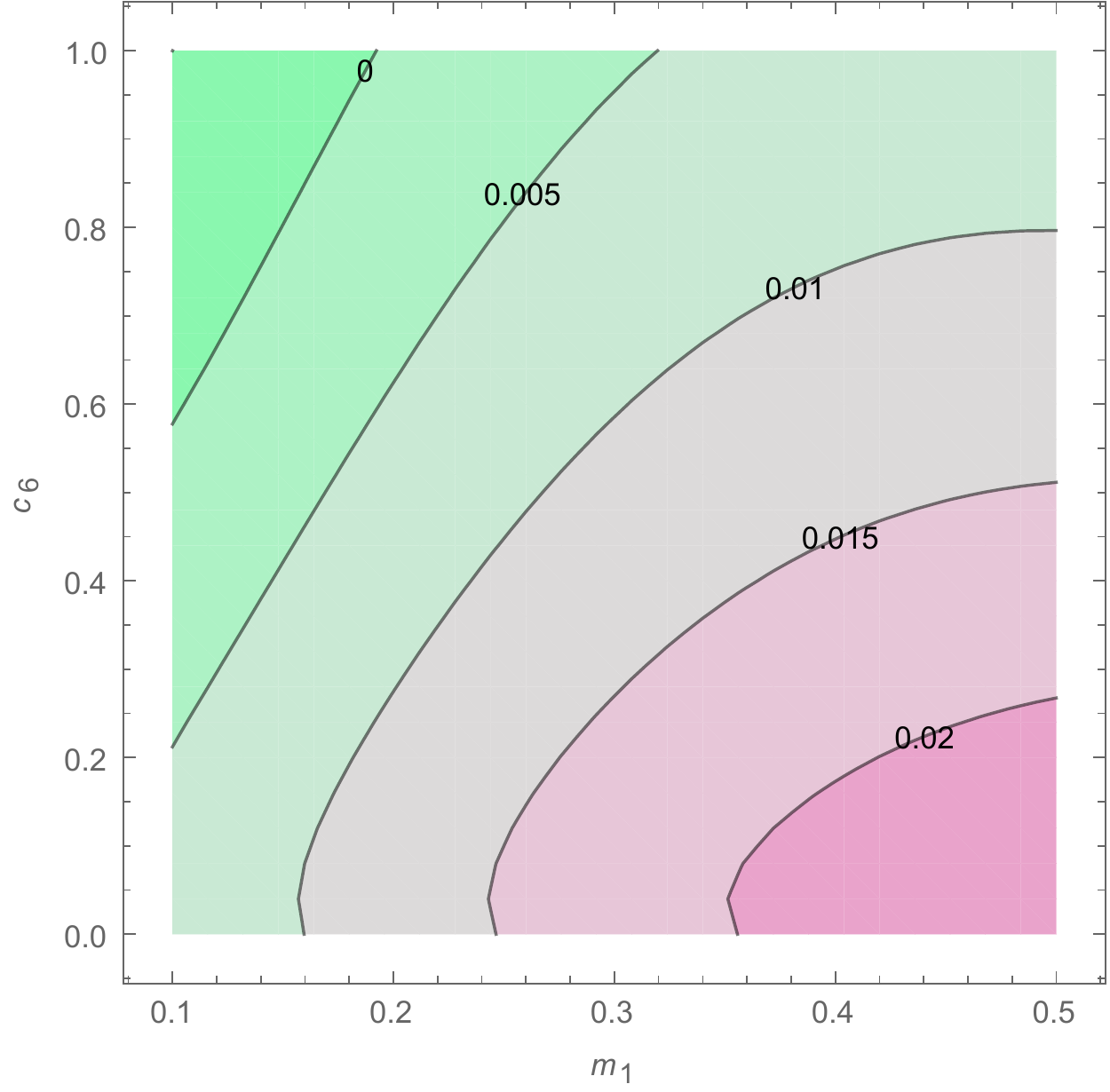}}}
  \end{center}
  \caption{Relative classical binding energy, $\delta_4$, in (a) the
    $(m_2,c_6)$ plane for $m_1=0.25$ and (b) the $(m_1,c_6)$ plane for
    $m_2=0.7$. 
  }
  \label{fig:delta4}
\end{figure}

\begin{figure}[!htp]
  \begin{center}
    \mbox{
      \subfloat[]{\includegraphics[width=0.49\linewidth]{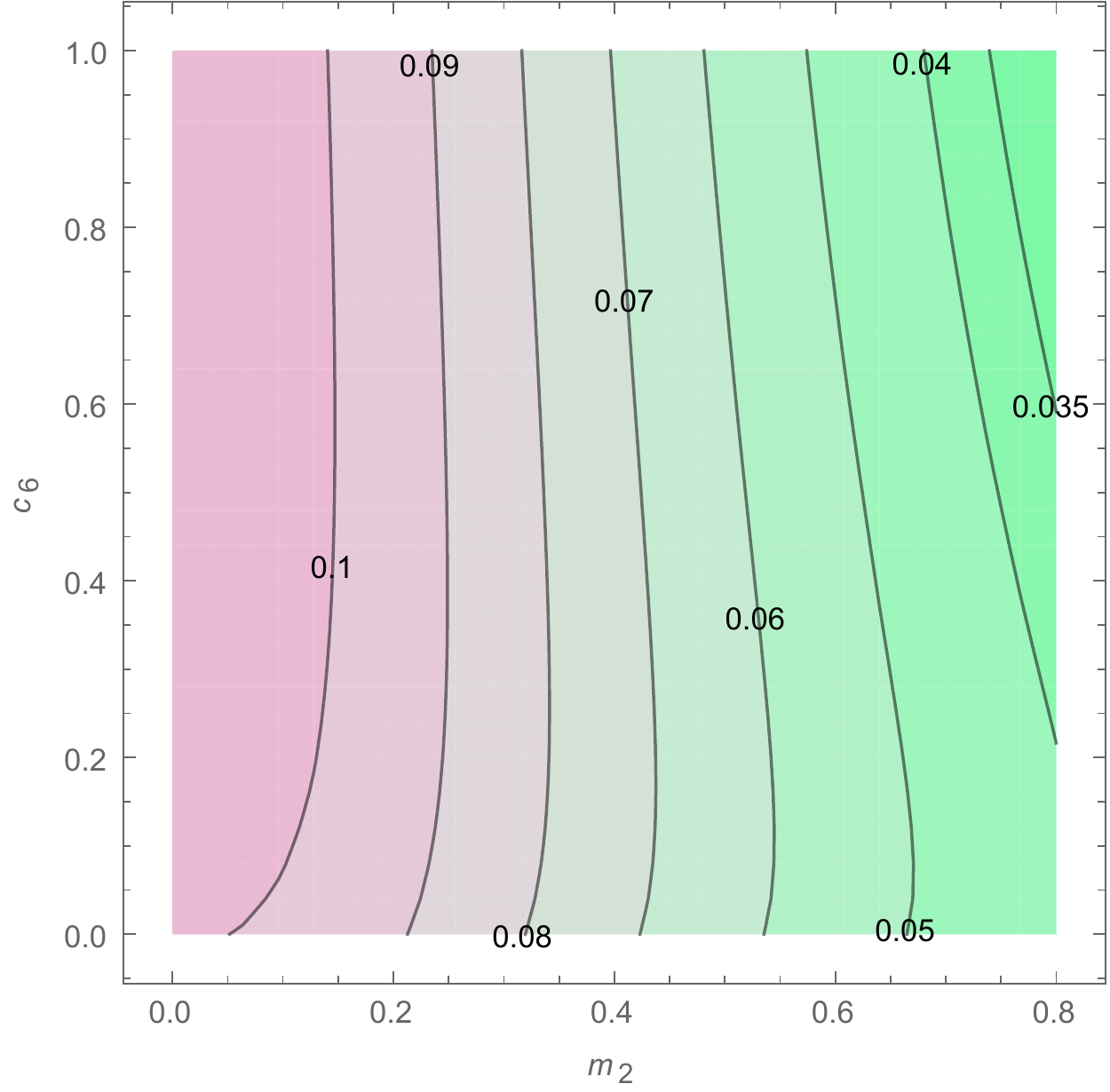}}
      \subfloat[]{\includegraphics[width=0.49\linewidth]{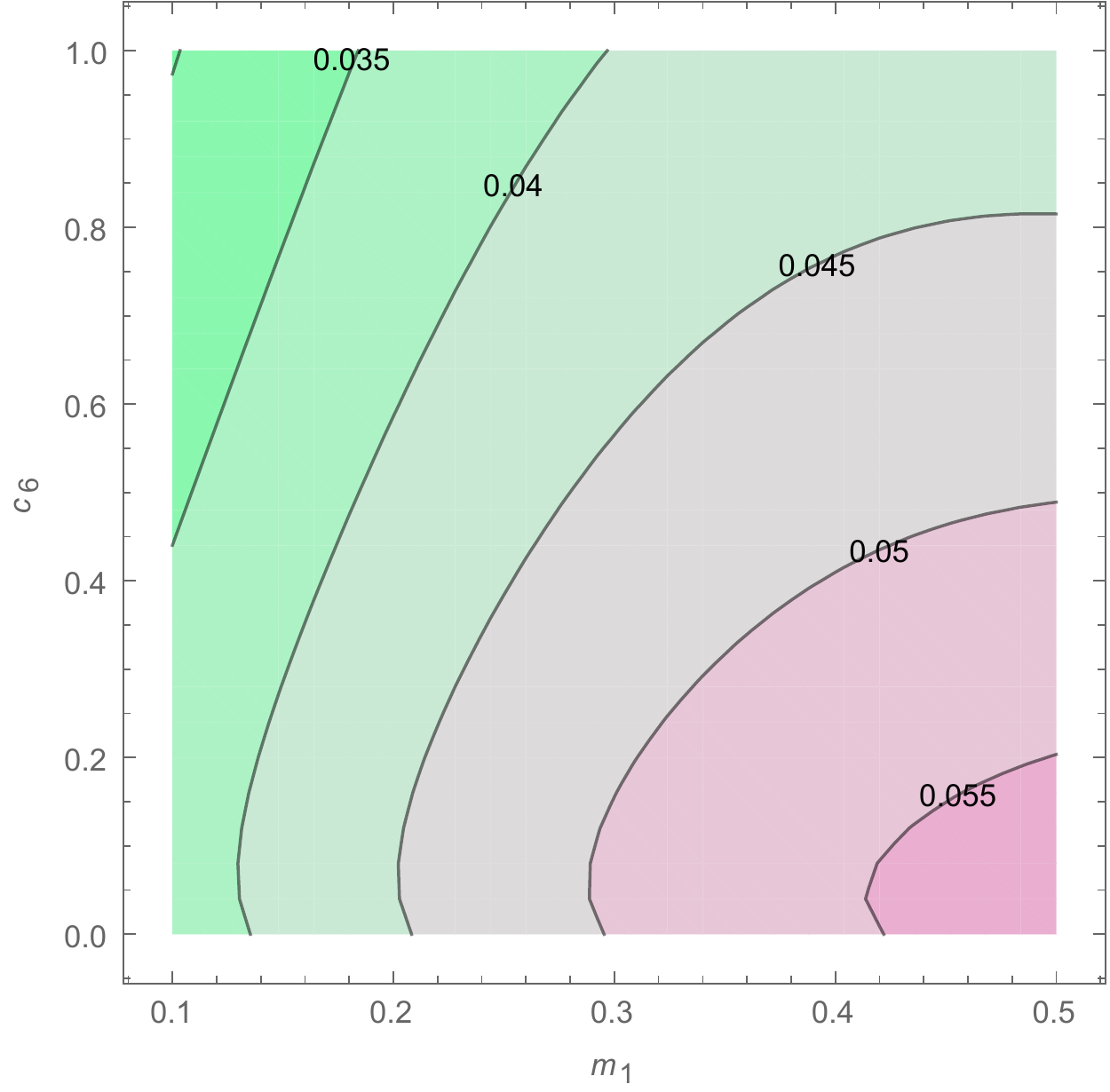}}}
  \end{center}
  \caption{Relative total binding energy, $\delta_4^{\rm tot}$, in (a)
    the $(m_2,c_6)$ plane for $m_1=0.25$ and (b) the $(m_1,c_6)$ plane
    for $m_2=0.7$. 
  }
  \label{fig:delta4tot}
\end{figure}

We will start by presenting the relative classical and total binding
energies in Figs.~\ref{fig:delta4} and \ref{fig:delta4tot},
respectively.
In these and the remaining figures containing contour plots in this
section, panel (a) shows the $(m_2,c_6)$-plane of parameter space for
fixed $m_1=0.25$ and (b) shows the $(m_1,c_6)$-plane for $m_2=0.7$. 

We see -- as expected from Ref.~\cite{Gudnason:2016cdo} --
that the binding energy decreases when $m_2$ is turned on and
monotonically decreases until $m_2=0.8$, which is the upper bound for 
$c_6=0$ for which the cubic symmetry of the 4-Skyrmion is still
retained.
Now we turn on the BPS-Skyrme term (whose coefficient is $c_6$) and we
can see that for $m_2=0$ it first increases the binding energy until a
plateau quickly is reached.
Interestingly, however, when $m_2=0.8$ the BPS-Skyrme term decreases
the binding energy and so much so that the classical binding energy in
Fig.~\ref{fig:delta4}a (Fig.~\ref{fig:delta4}b) turns negative in the
top-right (top-left) corner of the graph. 
We should warn the reader that since we are using the rational map
approximation for the 4-Skyrmion, the energy is expected to be
overestimated by about 1-2\% and since the 1-Skyrmion is an exact
numerical solution, this translates into a 1-2\% underestimation of
the binding energy.
Therefore the contour line marking $\delta_4=0$ actually corresponds
to a classical binding energy of about 1-2\%.
Shortly, we will estimate the systematic error by making a comparison
to the results of Ref.~\cite{Gudnason:2016cdo}.

Fig.~\ref{fig:delta4}b shows the $(m_1,c_6)$-plane of parameter space
for fixed $m_2=0.7$. It is a little surprising that the increase of
the pion mass $m_1$ leads to a slight increase in the binding energy.
The conclusions in both Refs.~\cite{Gudnason:2016mms} and
\cite{Gudnason:2016cdo} were that a larger pion mass was useful
because it was possible to reach smaller binding energies.
It is perfectly consistent, because the smaller binding energies were
reached by increasing the coefficient of the loosely bound potential,
$m_2$, and larger values of $m_1$ allow for larger values of $m_2$
before the cubic symmetry of the 4-Skyrmion is lost.
For fixed $m_2$, however, an increase in the pion mass is observed not
to help.

Taking the quantum contribution from spin-isospin quantization into
account gives us the total binding energy of the 4-Skyrmion, 
i.e.~$\delta_4^{\rm tot}$, shown in Fig.~\ref{fig:delta4tot}.
Exactly the same tendencies can be seen in the total binding energies
as in the classical binding energies.
The lowest total binding energy in Fig.~\ref{fig:delta4tot}a is
$\delta_4^{\rm tot}=3.03\%$ at $(m_2,c_6)=(0.8,1)$ and marginally less
at $(m_1,c_6)=(0.1,1)$, see Figs.~\ref{fig:delta4tot}a and
\ref{fig:delta4tot}b, respectively. 
Recall however, that due to the systematic error in using the rational
map approximation, the true total binding energy is expected to be
about 1-2\% higher.

\begin{figure}[!htp]
  \begin{center}
    \mbox{
      \subfloat[]{\includegraphics[width=0.49\linewidth]{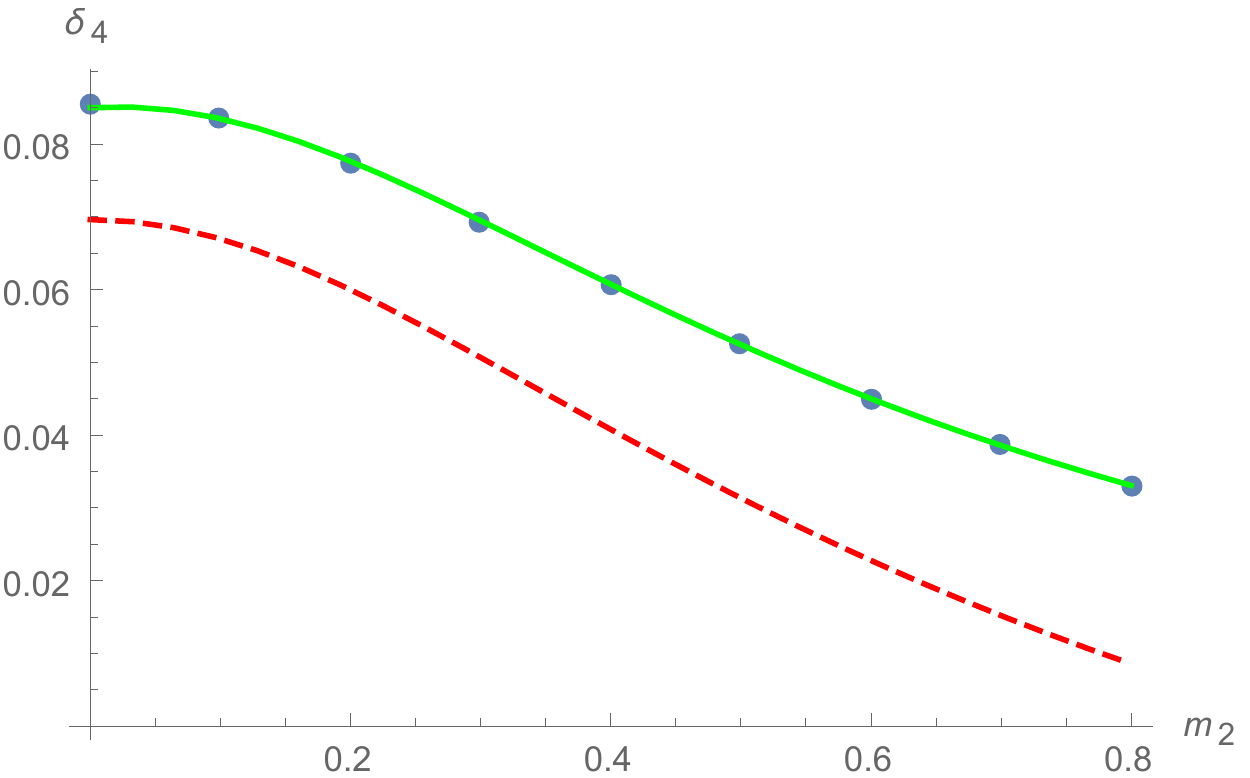}}
      \subfloat[]{\includegraphics[width=0.49\linewidth]{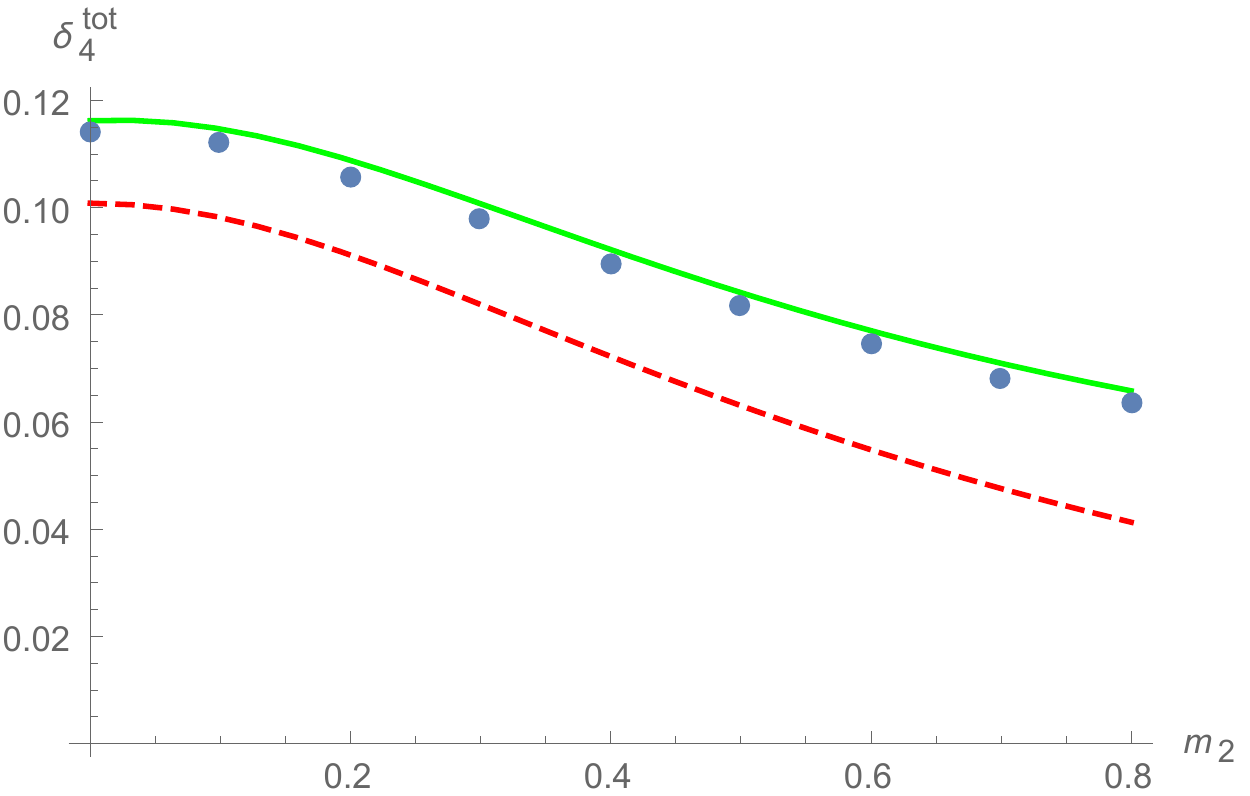}}}
  \end{center}
  \caption{Relative (a) classical and (b) total binding energy,
    $\delta_4$ and $\delta_4^{\rm tot}$, as functions of $m_2$ for
    $m_1=0.25$. 
    The blue circles are full PDE calculations from
    Ref.~\cite{Gudnason:2016cdo}, the red dashed line is the
    calculation using the rational map approximation and finally the
    green solid line is the rational map approximation with a
    correction for systematic error, see
    Eq.~\eqref{eq:delta4corrected}. 
  }
  \label{fig:delta4sys}
\end{figure}

Let us now address the systematic error in the binding energy due to
the rational map approximation.
We take the binding energies calculated from the full PDE solutions in
Ref.~\cite{Gudnason:2016cdo} and compare them to the slice in our
parameter space where $c_6=0$.
Using a linear function in $m_2$, we will fit the difference between
the full PDE solutions and the rational map approximations
\beq
\delta_4^{\rm corrected} = \delta_4 + a + b m_2,
\label{eq:delta4corrected}
\eeq
where the constants are determined as $a=0.01538$ and $b=0.01142$.
We also find that the dependence of $m_1$ of the difference is an
order of magnitude smaller than that of $m_2$ and so we will ignore it
here. 
Fig.~\ref{fig:delta4sys} shows the full PDE calculation as blue
circles, the rational map approximation as the red dashed line and
finally the corrected relative binding energy of
Eq.~\eqref{eq:delta4corrected} as the green solid line.
The linear fit tells us that the energy of the 4-Skyrmion is
\emph{indeed} overestimated, which in turn, underestimates the binding
energy by 1.5\% for $m_2=0$ and by 2.5\% for $m_2=0.8$. 

We used the classical binding energy to fit the correction for the
systematic error of using the rational map approximation and as we can
see in Fig.~\ref{fig:delta4sys}a, the linear correction works very
well.
Using the same correction, we can see in Fig.~\ref{fig:delta4sys}b,
that the corrected rational map approximation \emph{overestimates} the
true binding energy by a little.
Nevertheless with the correction for the systematic error, the
relative classical binding energy matches the full PDE calculation
within 0.0042-0.065\%, whereas the relative total binding energies
only matches the PDEs within 0.18-0.29\%. 

\begin{figure}[!htp]
  \begin{center}
    \mbox{
      \subfloat[]{\includegraphics[width=0.49\linewidth]{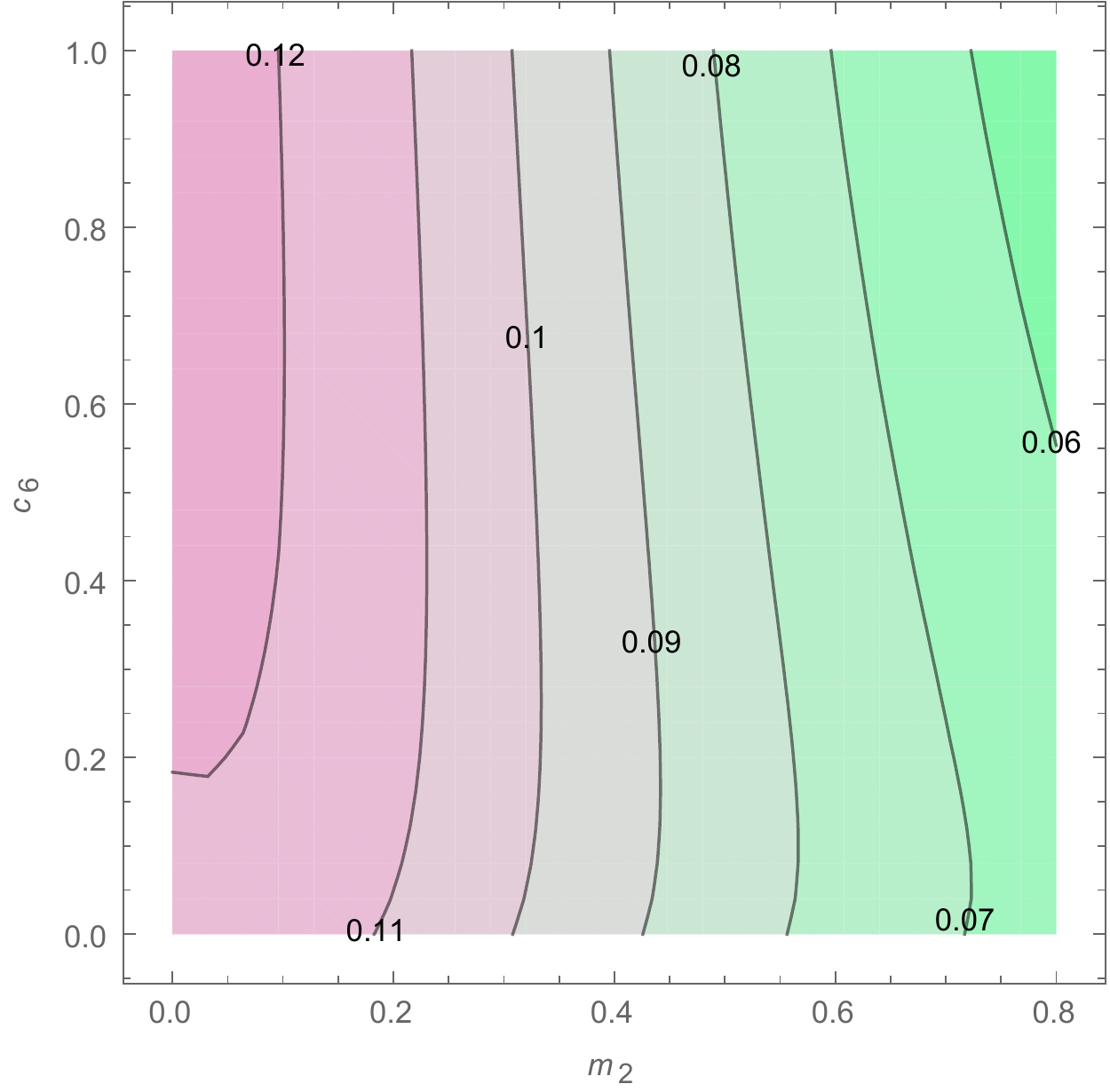}}
      \subfloat[]{\includegraphics[width=0.49\linewidth]{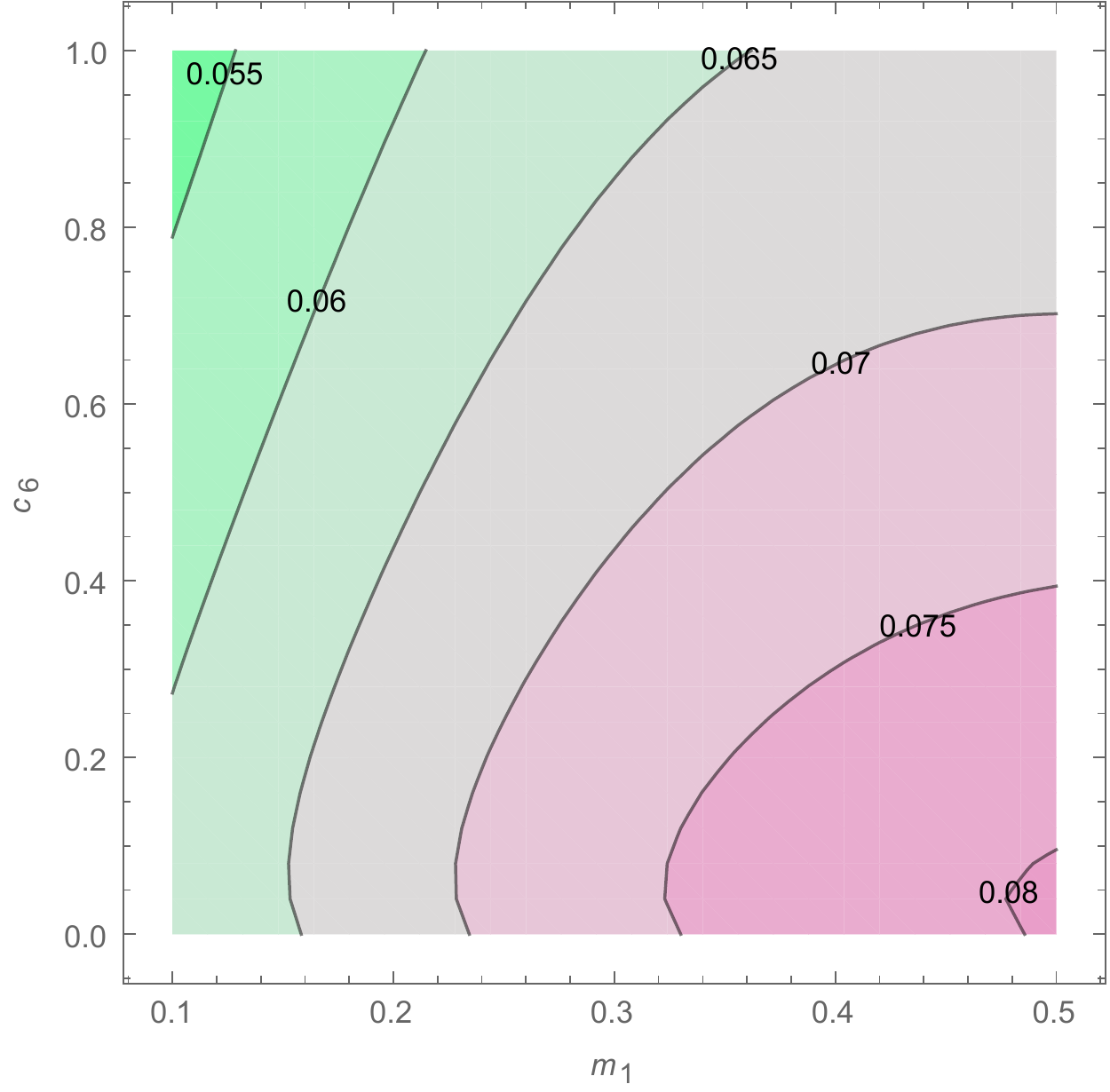}}}
  \end{center}
  \caption{Relative total binding energy with a linear correction of
    the systematic error due to the rational map approximation,
    $\delta_4^{\rm corrected}$, in (a) the $(m_2,c_6)$ plane for
    $m_1=0.25$ and (b) the $(m_1,c_6)$ plane for $m_2=0.7$. 
  }
  \label{fig:delta4totoffset}
\end{figure}

Finally, we will correct the systematic error of using the rational
map approximation based on the above fit and show the relative total
binding energy in Fig.~\ref{fig:delta4totoffset}.

Interestingly, we can see that the BPS-Skyrme term helps decreasing
the binding energy for large $m_2$ and at the top-right corner of
Fig.~\ref{fig:delta4totoffset}a we have about 5.5\% binding energy and
in the top-left corner of Fig.~\ref{fig:delta4totoffset}b it is about
5.3\%.
Compared to turning off the BPS-Skyrme term ($c_6=0$) the latter two
values read 6.6\% and 6.1\%, respectively.
With the results of Ref.~\cite{Gudnason:2016cdo} in mind, we expect
that instead of decreasing the pion mass, increasing it and in the
same time increasing also $m_2$, but beyond the values explored here,
we will be able to reach even lower binding energies.

\subsection{Calibration}

\begin{figure}[!htp]
  \begin{center}
    \mbox{
      \subfloat[]{\includegraphics[width=0.49\linewidth]{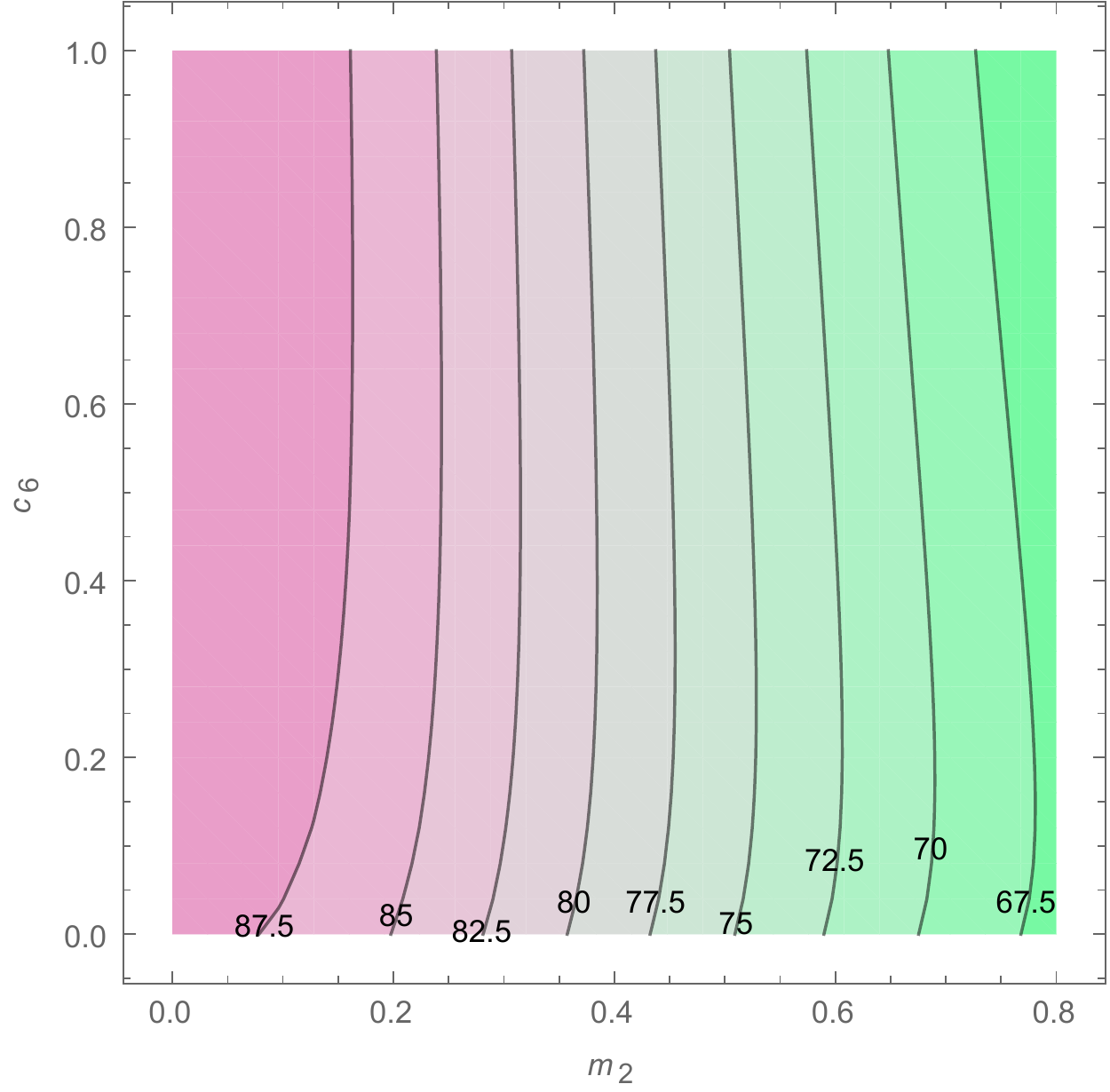}}
      \subfloat[]{\includegraphics[width=0.49\linewidth]{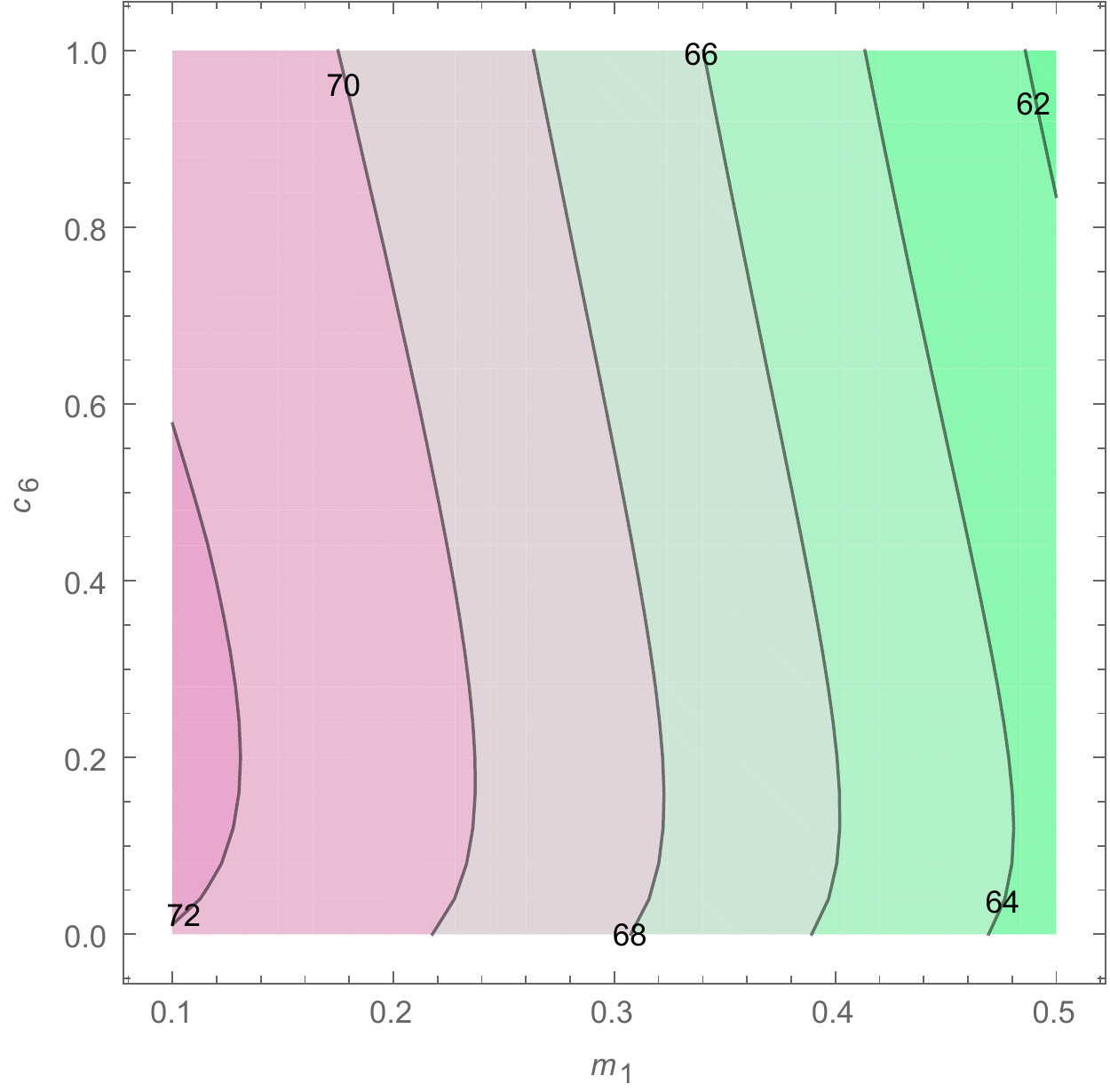}}}
  \end{center}
  \caption{Pion decay constant, $f_\pi$ [MeV], in (a) the $(m_2,c_6)$
    plane for $m_1=0.25$ and (b) the $(m_1,c_6)$ plane for $m_2=0.7$. 
  }
  \label{fig:fpi}
\end{figure}

In the calculation of the quantum contribution, which is an ingredient
in the binding energies discussed above, the calibration of the model
of the 4-Skyrmion to ${}^4$He, has been used, see
Eq.~\eqref{eq:fpi_e_calib}.
Although the relative binding energy does not depend on the pion decay
constant, $f_\pi$, other observables like the pion mass do depend on
it.
In Fig.~\ref{fig:fpi} is shown the pion decay constant in the explored
parameter space.
We can see that increasing either the pion mass, $m_1$, or the
coefficient of the loosely bound potential, $m_2$, with the other one
held fixed, decreases the already too small pion decay constant.
For large $m_2\sim 0.8$, the pion decay constant in the model comes
out as low as 67.5 MeV, compared to the experimental value of about
186 MeV; a factor of 3 too low. 

\begin{figure}[!htp]
  \begin{center}
    \mbox{
      \subfloat[]{\includegraphics[width=0.49\linewidth]{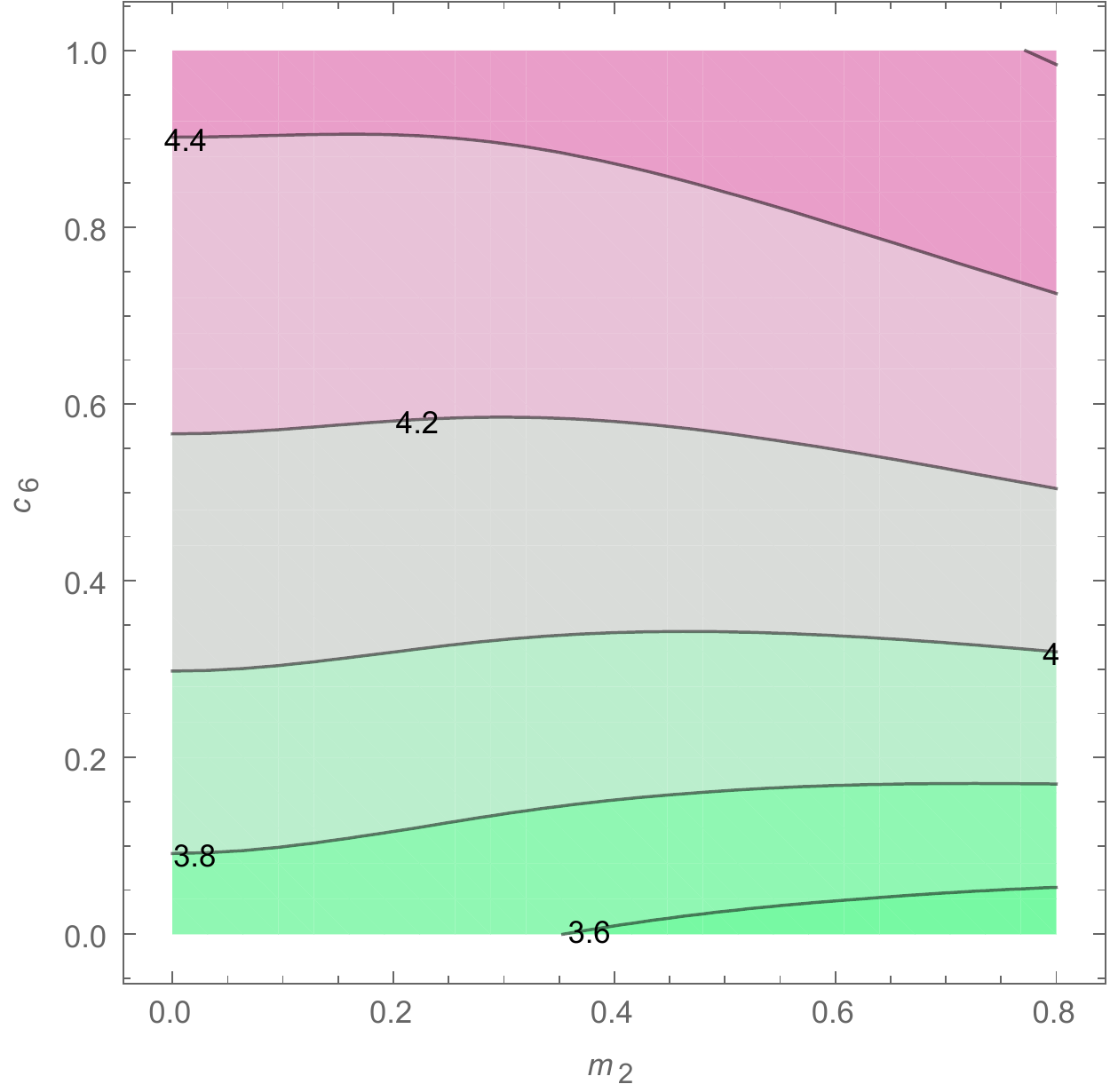}}
      \subfloat[]{\includegraphics[width=0.49\linewidth]{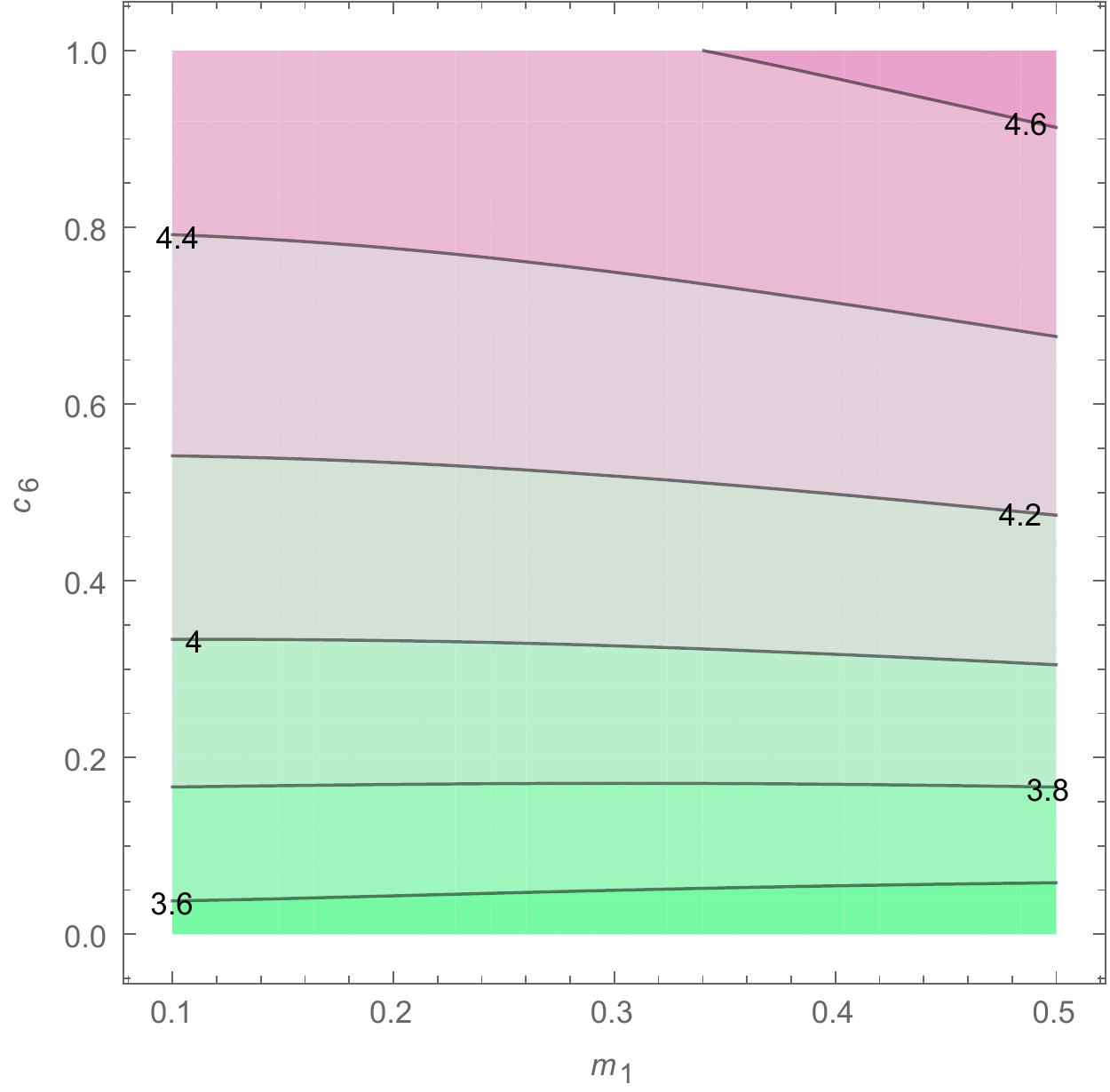}}}
  \end{center}
  \caption{Skyrme term coefficient, $e$, in (a) the $(m_2,c_6)$
    plane for $m_1=0.25$ and (b) the $(m_1,c_6)$ plane for $m_2=0.7$. 
  }
  \label{fig:e}
\end{figure}

The next parameter is the Skyrme-term coefficient $e$, which to the
best of our knowledge is not experimentally determined.
The calibration determines its value, which is shown in
Fig.~\ref{fig:e}.

The general tendency of turning on the BPS-Skyrme term is an increase
in $e$.
Let us recall that the prefactor of the quantum correction to the
energy of the 1-Skyrmion has an $e^4$ relative to its classical
value. This naively implies that larger $e$ leads to larger binding
energies.
As we know from Eq.~\eqref{eq:gAtilde}, a larger $e$ will also have
the effect of decreasing the axial coupling, which we will see
shortly is a desired feature.

\subsection{Mass spectrum}

\begin{figure}[!htp]
  \begin{center}
    \mbox{
      \subfloat[]{\includegraphics[width=0.49\linewidth]{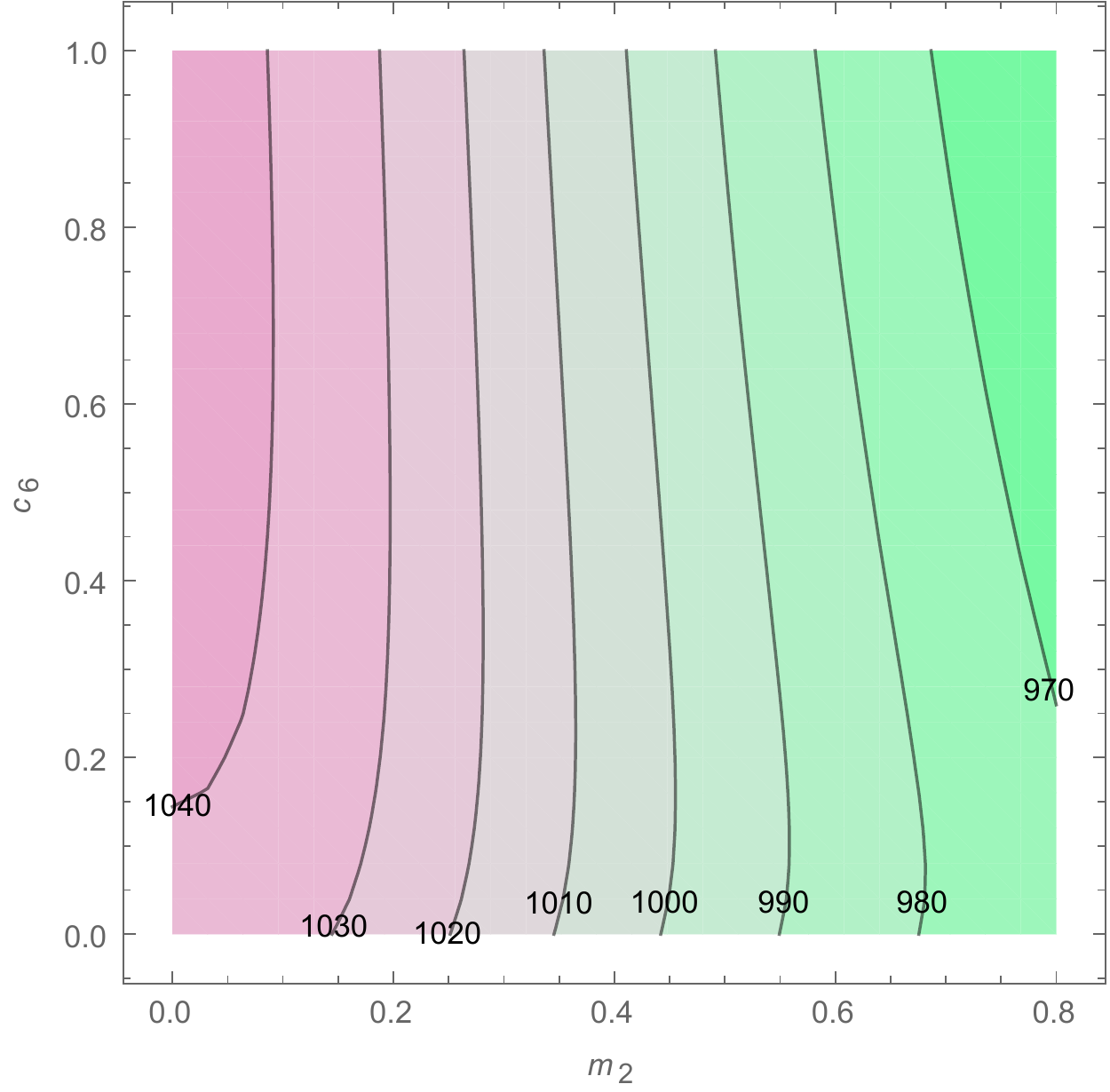}}
      \subfloat[]{\includegraphics[width=0.49\linewidth]{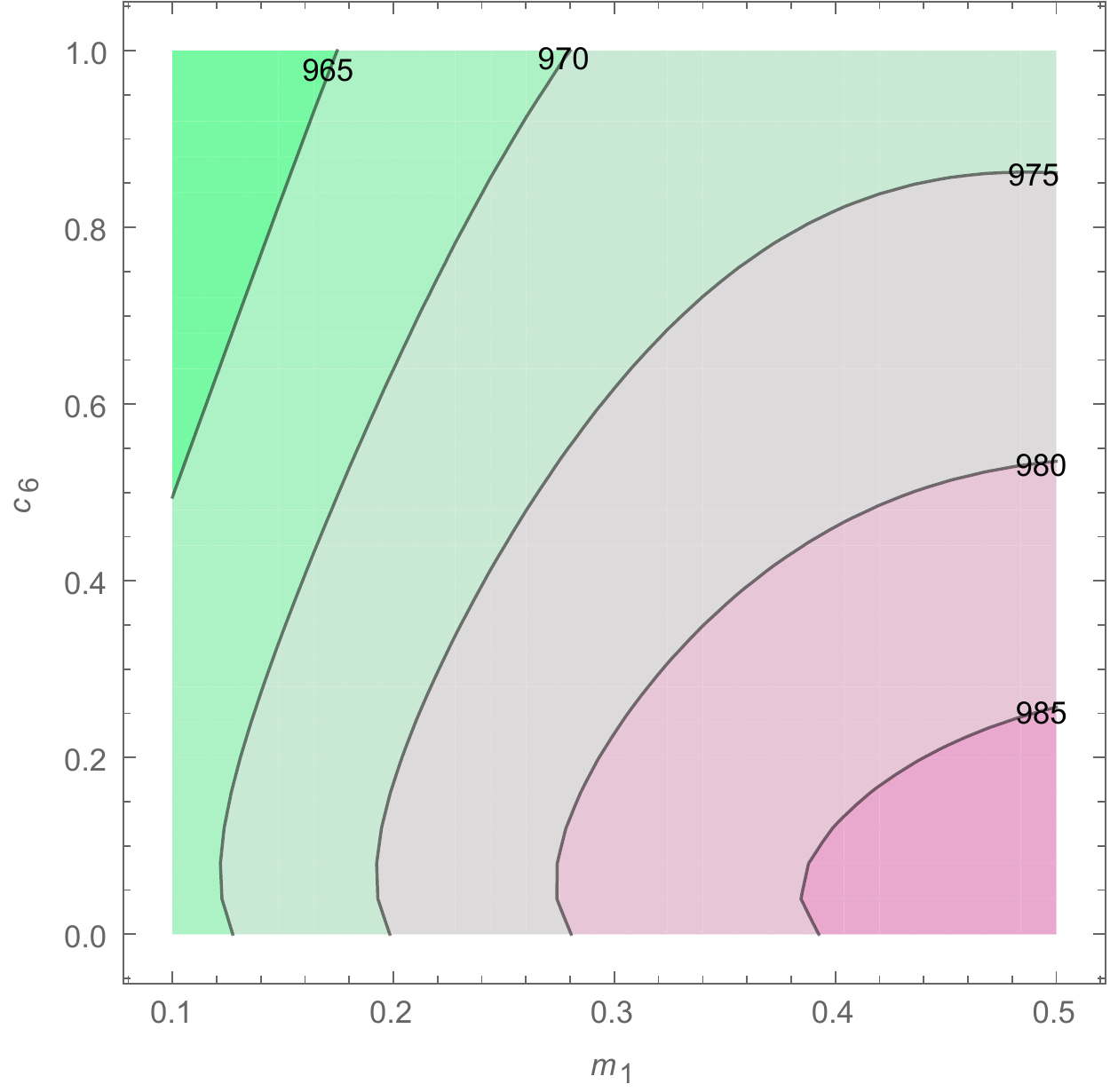}}}
  \end{center}
  \caption{Nucleon mass, $\tilde{m}_N$ [MeV], in (a) the $(m_2,c_6)$
    plane for $m_1=0.25$ and (b) the $(m_1,c_6)$ plane for $m_2=0.7$. 
  }
  \label{fig:mN}
\end{figure}

We are now ready to calculate the physical spectrum, which we limit to
the nucleon mass, the mass of the Delta resonance and the pion mass.
The nucleon mass is shown in Fig.~\ref{fig:mN}.
Recall that we calibrate the model by setting the mass of the
4-Skyrmion equal to the mass of Helium-4.

We can see from Fig.~\ref{fig:mN}a that increasing the loosely bound 
potential, decreases the nucleon mass, but yet not enough to reach its
experimentally measured value around 939 MeV. 
Similarly to the relative binding energy, when $m_2=0$ the BPS-Skyrme
term initially increases the nucleon mass until a plateau quickly is
reached. However, when $m_2$ is large the BPS-Skyrme term decreases
the value of the nucleon mass.
It is interesting to see from Fig.~\ref{fig:mN}b, that a smaller pion
mass gives rise to a smaller nucleon mass. 
The part of parameter space we considered here generally overestimates
the nucleon mass.

\begin{figure}[!htp]
  \begin{center}
    \mbox{
      \subfloat[]{\includegraphics[width=0.49\linewidth]{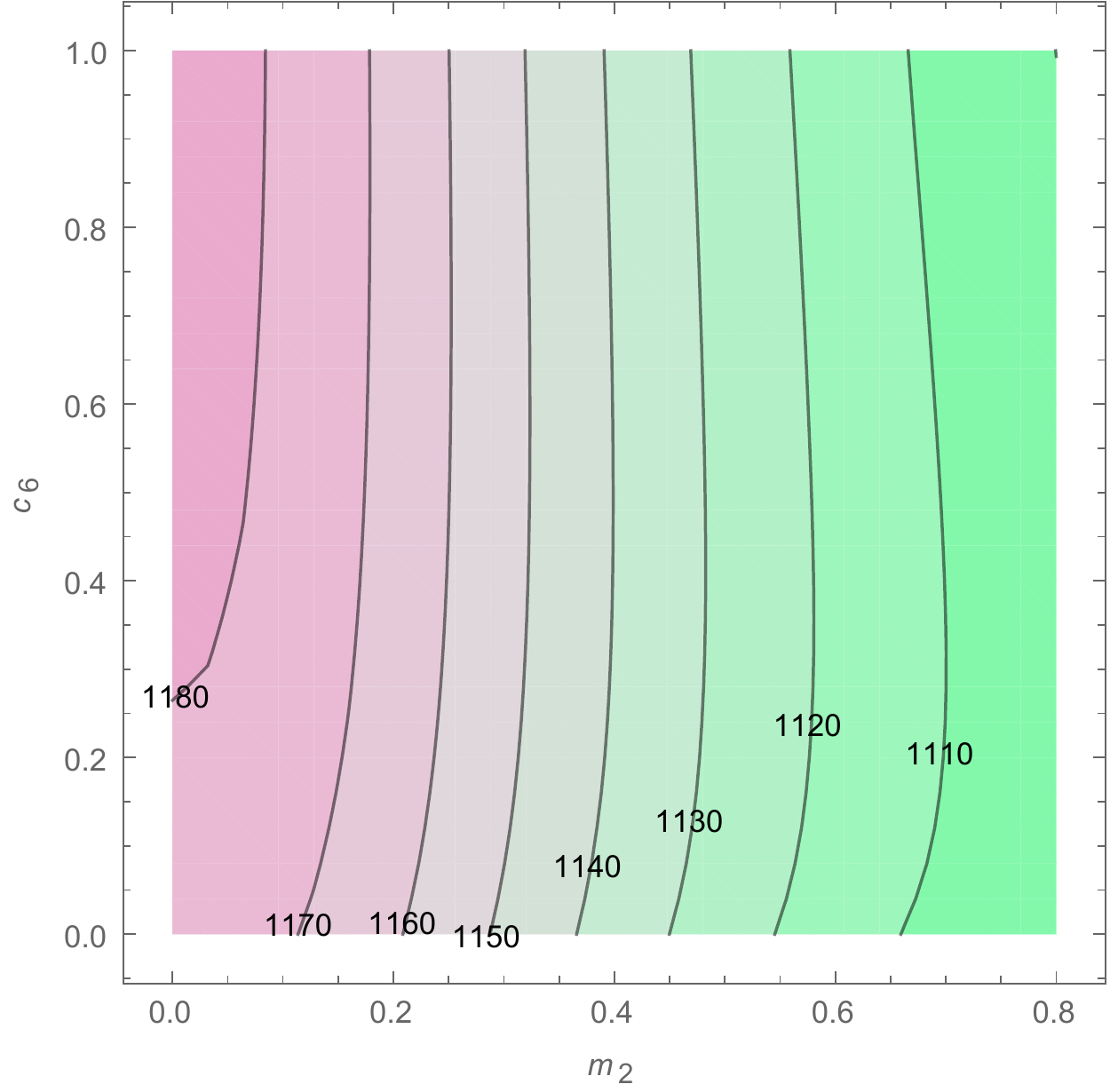}}
      \subfloat[]{\includegraphics[width=0.49\linewidth]{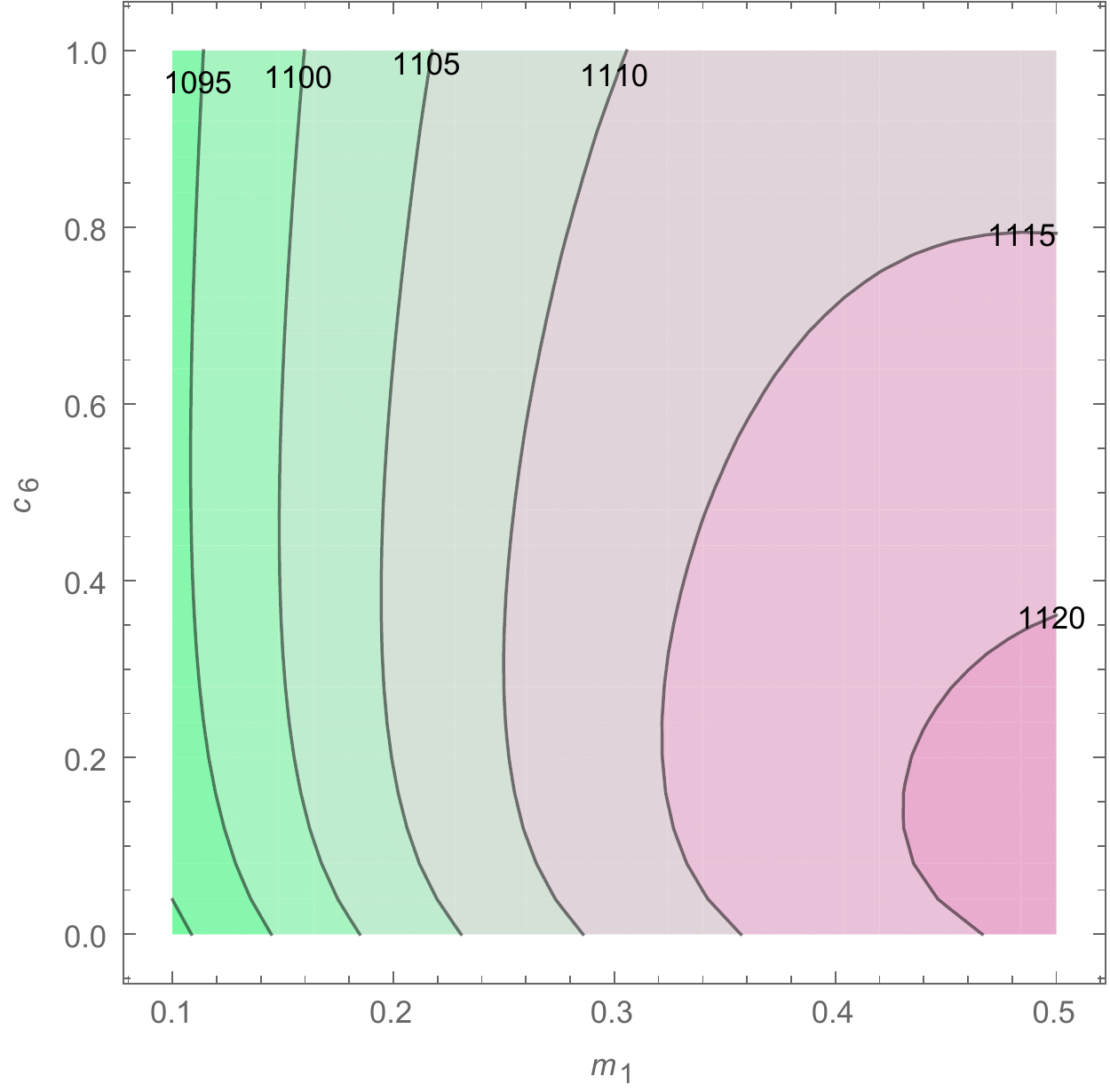}}}
  \end{center}
  \caption{Mass of the $\Delta$ resonance, $\tilde{m}_\Delta$ [MeV],
    in (a) the $(m_2,c_6)$ plane for $m_1=0.25$ and (b) the
    $(m_1,c_6)$ plane for $m_2=0.7$. 
  }
  \label{fig:mDelta}
\end{figure}

We now turn to the mass of the Delta resonance, which is shown in
Fig.~\ref{fig:mDelta}. It is well known that the Skyrme model
generally underestimates it, and our flavor of the Skyrme model is no
exception in this part of the parameter space.

The tendency of improvement of the binding energy and nucleon mass for
larger values of the loosely bound potential ($m_2$) leads, however,
to an exacerbation of the mass of the Delta resonance being too
small.

\begin{figure}[!htp]
  \begin{center}
    \mbox{
      \subfloat[]{\includegraphics[width=0.49\linewidth]{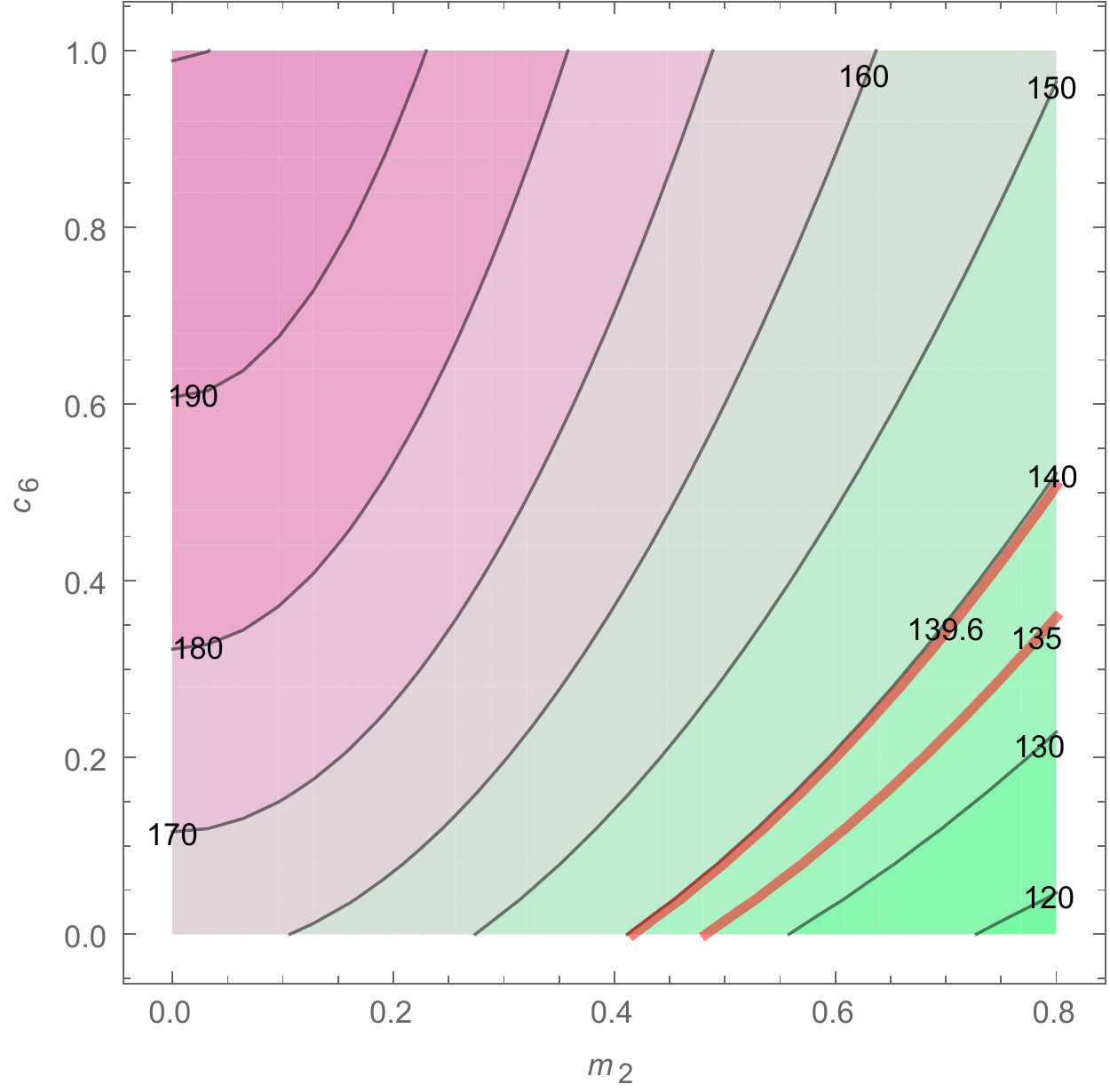}}
      \subfloat[]{\includegraphics[width=0.49\linewidth]{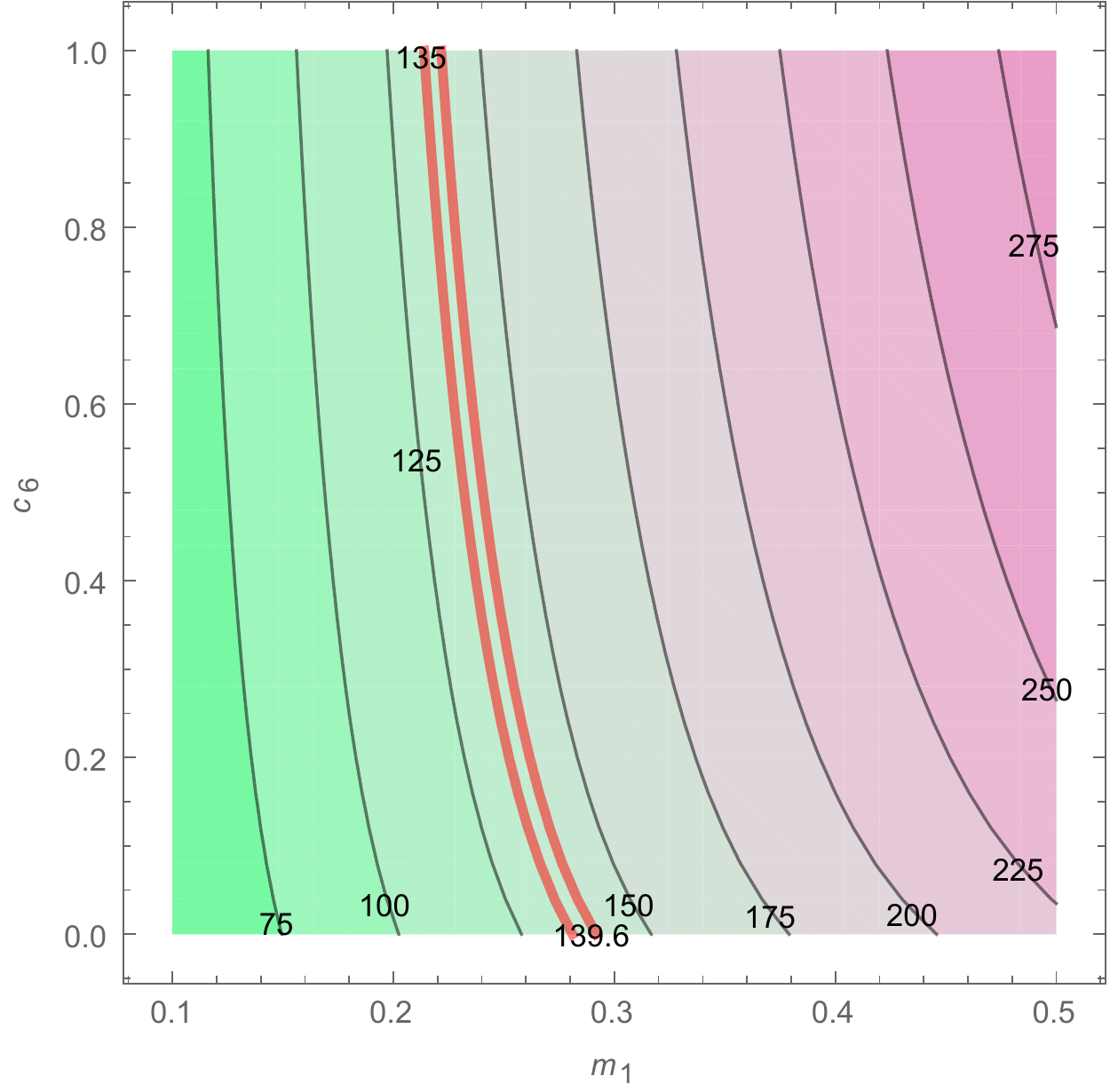}}}
  \end{center}
  \caption{Pion mass, $\tilde{m}_\pi$ [MeV], in (a) the $(m_2,c_6)$
    plane for $m_1=0.25$ and (b) the $(m_1,c_6)$ plane for $m_2=0.7$.
    The two bold red lines indicate the physical values of the pion
    masses. 
  }
  \label{fig:mpi}
\end{figure}

The value of the pion mass is shown in Fig.~\ref{fig:mpi}.
This is the first observable hitting spot-on the experimentally
measured value.
Of course the charged pions in Nature are slightly heavier than the
neutral one, but since we leave chiral symmetry unbroken, all 3 pions
are mass degenerate in our model.
We can see that in order to hit the right experimental value and in
the same time reduce the binding energy by turning on the BPS-Skyrme
term, we need to reduce the value of the pion mass, $m_1$.
However, the BPS-Skyrme term is expected to bind the constituents of
the 4-Skyrmion more tightly, whereas the loosely bound potential
repels them from each other.
Therefore, we expect that it will be possible to increase the value of
$m_2$ for medium/large values of $c_6$ and so the tendency of the
model matching the experimental value of the pion mass is right on
track.

Let us, however, remark that the pion decay constant is about a factor
of 3 too small compared with experiment and in the modern point of
view in the Skyrme model we accept this fact as the pion decay
constant in the model being simply a renormalized value in the
effective field theory and in the baryon medium.

\subsection{Proton charge radius}

\begin{figure}[!htp]
  \begin{center}
    \mbox{
      \subfloat[]{\includegraphics[width=0.49\linewidth]{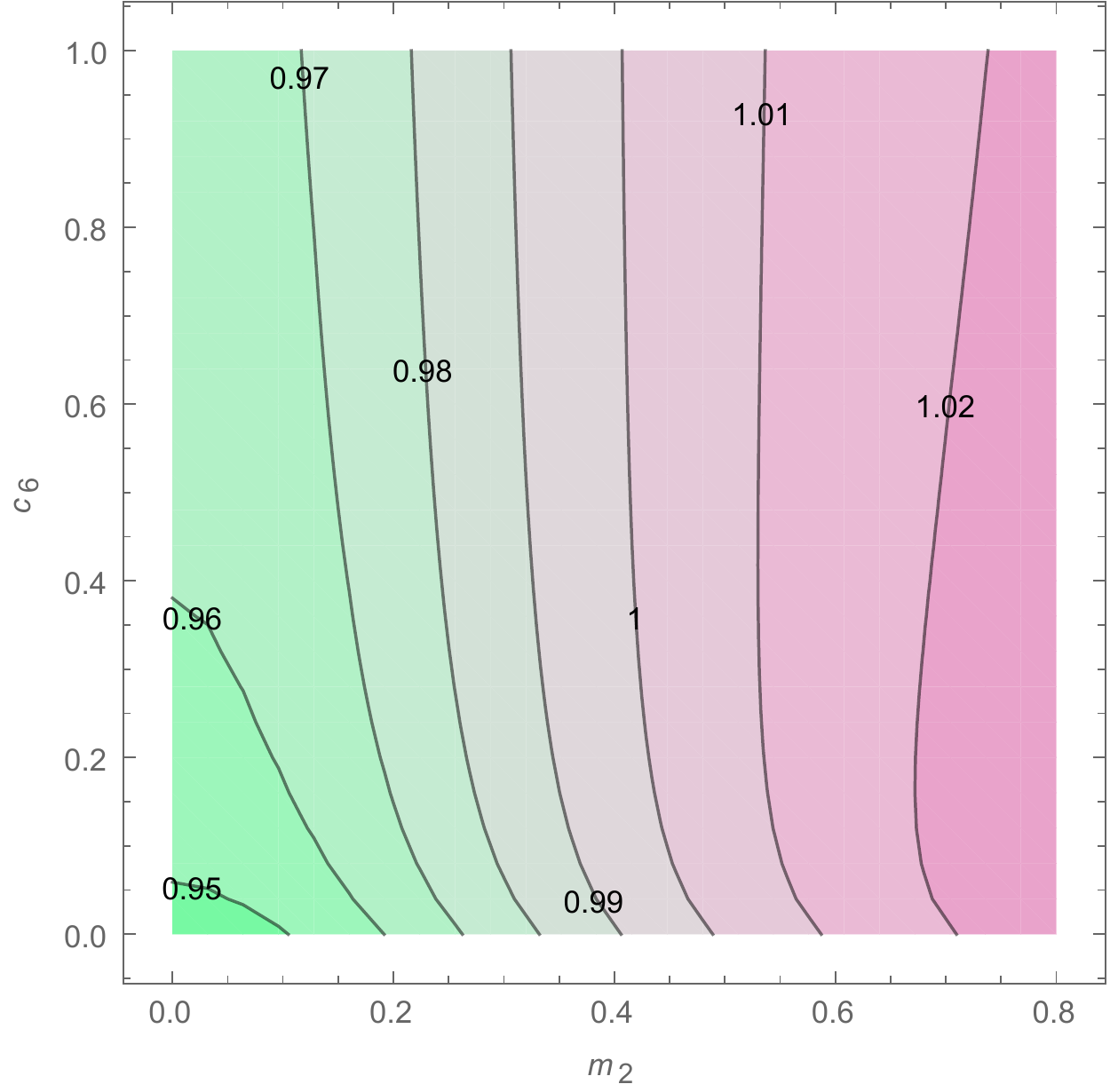}}
      \subfloat[]{\includegraphics[width=0.49\linewidth]{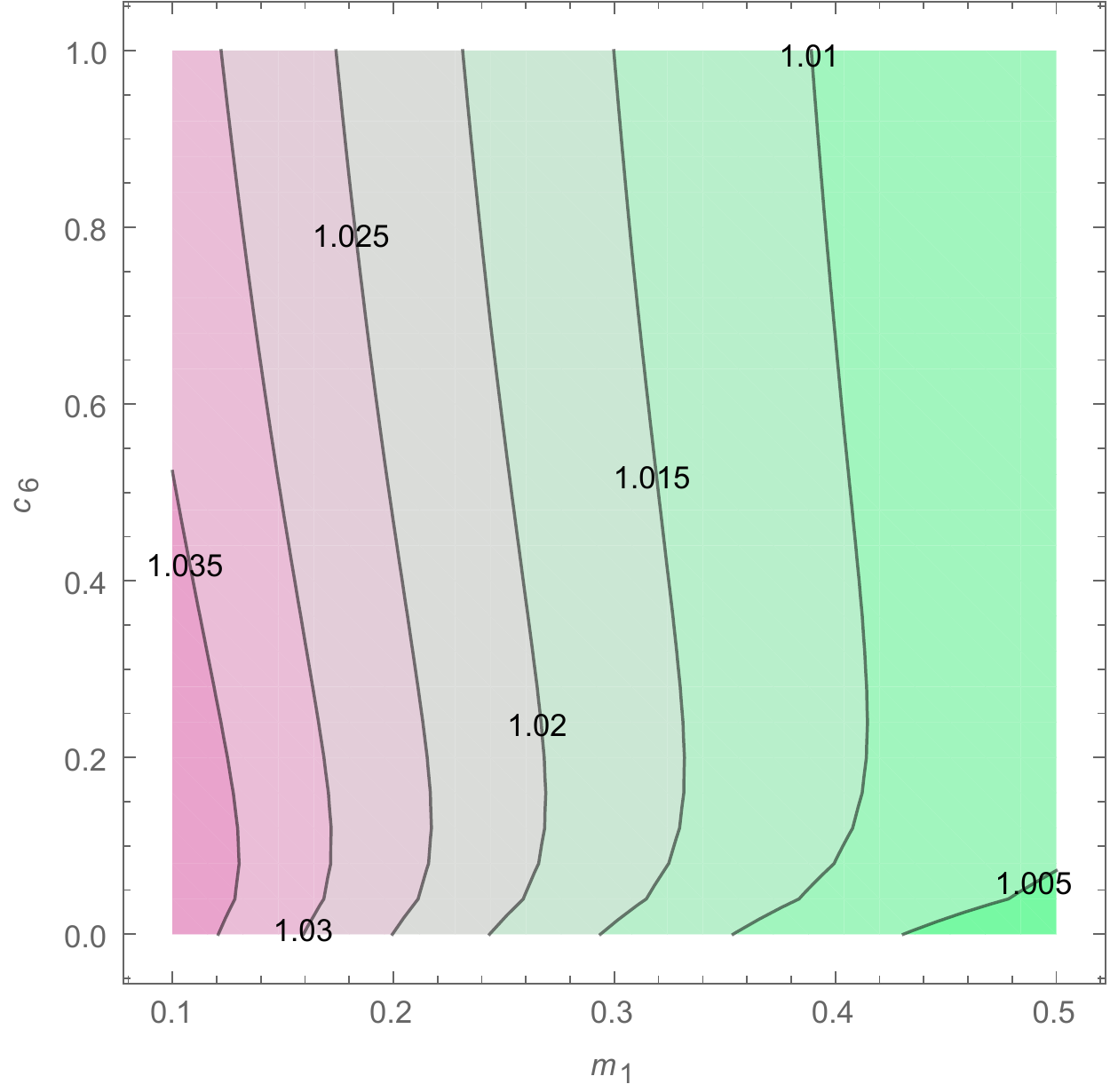}}}
  \end{center}
  \caption{Baryon charge radius of the proton,
    $r_1$ [fm], in (a) the $(m_2,c_6)$ plane for 
    $m_1=0.25$ and (b) the $(m_1,c_6)$ plane for $m_2=0.7$. 
  }
  \label{fig:rpB}
\end{figure}

\begin{figure}[!htp]
  \begin{center}
    \mbox{
      \subfloat[]{\includegraphics[width=0.49\linewidth]{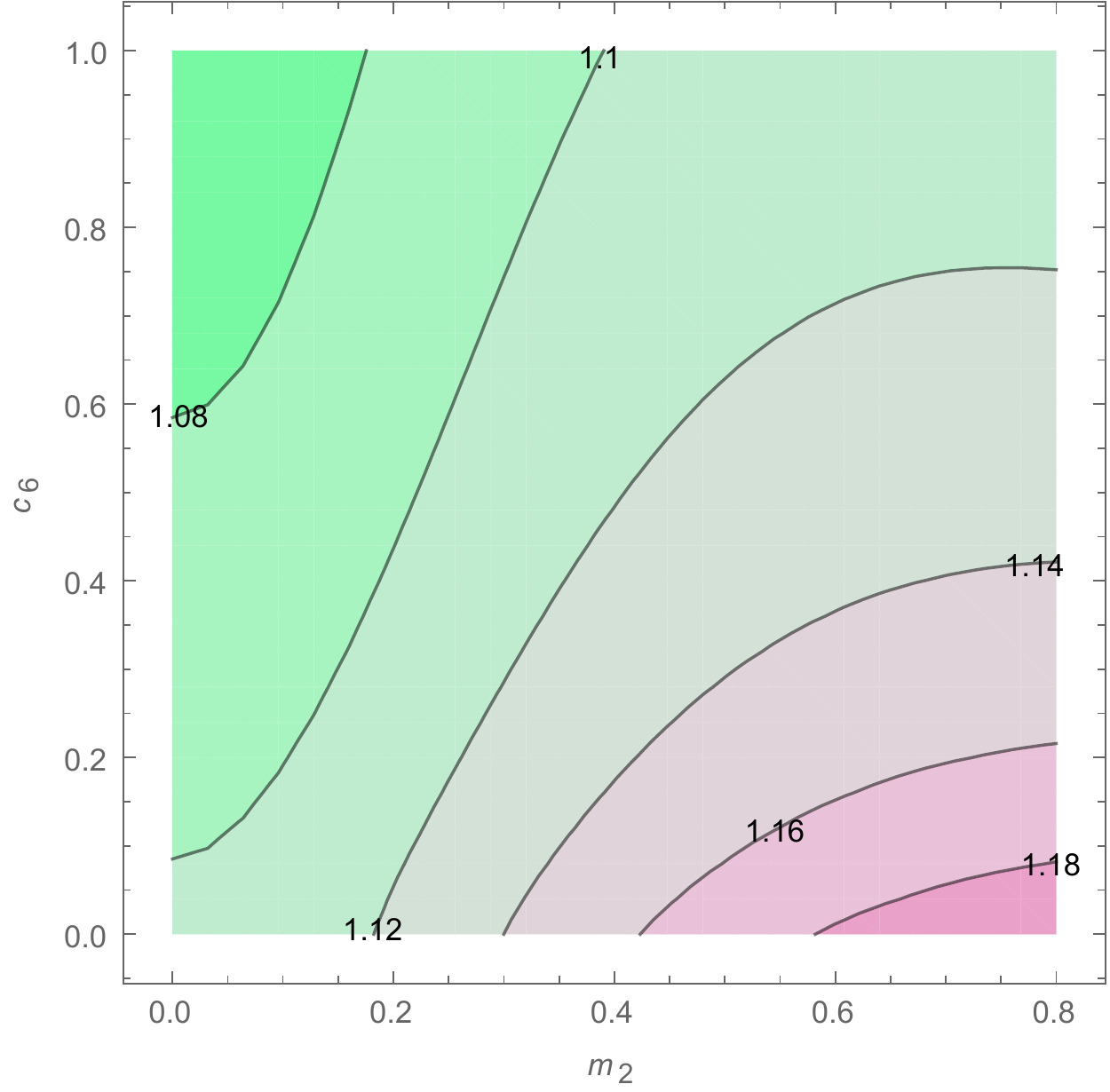}}
      \subfloat[]{\includegraphics[width=0.49\linewidth]{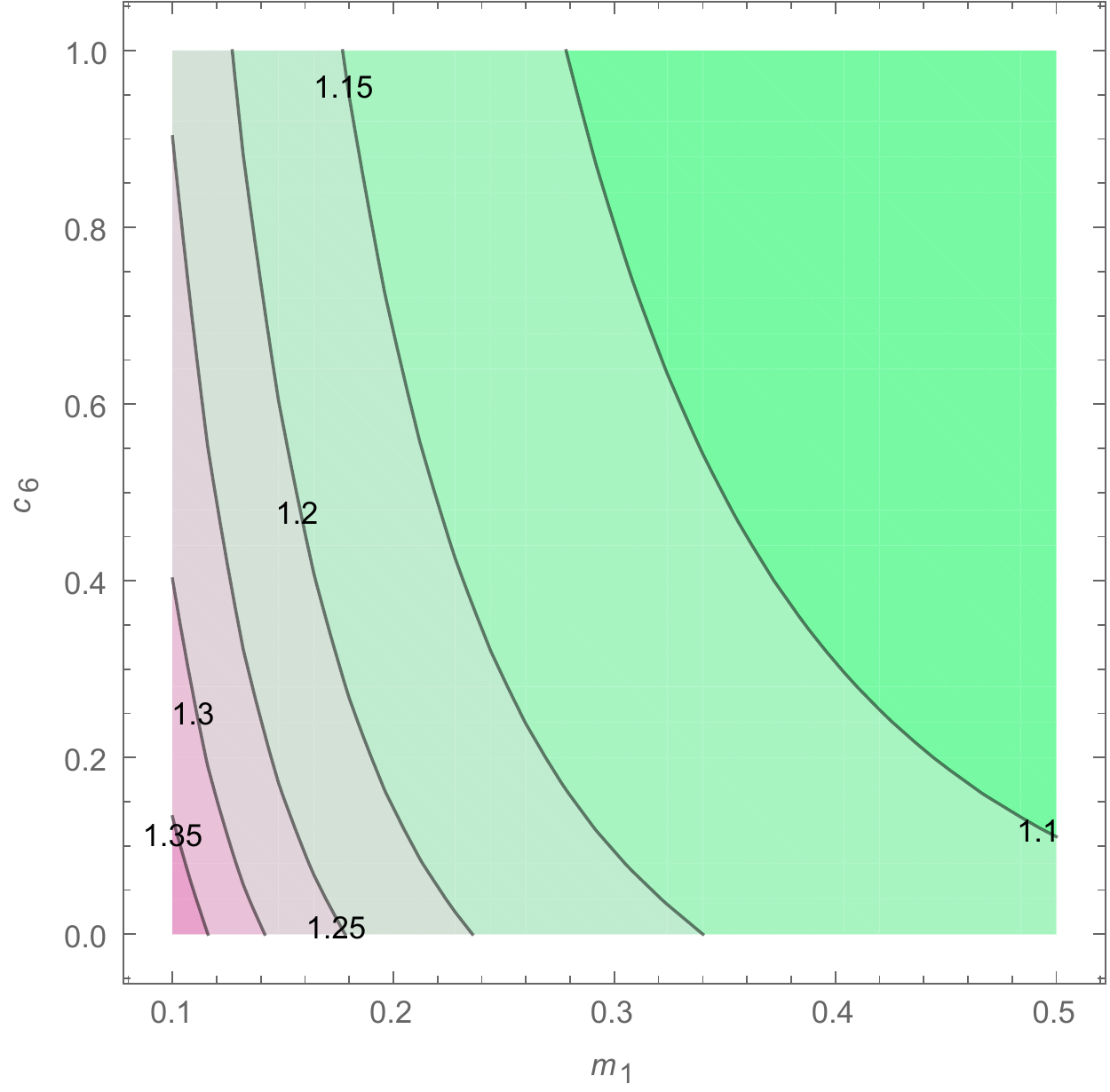}}}
  \end{center}
  \caption{Electric charge radius of the proton,
    $r_{E,1}$ [fm], in (a) the $(m_2,c_6)$ plane for 
    $m_1=0.25$ and (b) the $(m_1,c_6)$ plane for $m_2=0.7$. 
  }
  \label{fig:rpE}
\end{figure}

We now turn to the proton charge radius.
For comparison, we calculate both the baryon charge radius and the
electric charge radius, following Ref.~\cite{Adkins:1983ya}, but
generalized to include the BPS-Skyrme term, see Eq.~\eqref{eq:rE1}. 
The baryon charge radius, defined in Eq.~\eqref{eq:rB1}, is shown in
Fig.~\ref{fig:rpB} and the electric charge radius is shown in
Fig.~\ref{fig:rpE}. 

The model generally overestimates the proton charge radius.
If we begin with Fig.~\ref{fig:rpB}, we can see that the loosely bound
potential -- although lowering the binding energy -- increases the
baryon charge radius.
This can readily be understood as follows. The loosely bound
potential, like the mass term, decreases the size of the 4-Skyrmion.
However, since it is calibrated against the physical size of Helium-4,
this implies a prolongation of the length scales.
Because the 1-Skyrmion is more compact, it does not shrink as much as
the 4-Skyrmion in the presence of the loosely bound potential and with
the prolonged length scales, its size increases.
This yields a larger proton radius for larger values of the
coefficient of the loosely bound potential, $m_2$.
Interestingly, and counterintuitively, an increase of the pion mass
for fixed larger $m_2$, see Fig.~\ref{fig:rpB}, yields a
\emph{smaller} baryon charge radius.
The effect is however relatively small compared to that of changing
$m_2$. 

Since the proton is an electrically charged object, it has a
well-defined electric charge radius, which is experimentally much more 
easily accessible than the baryon charge radius.
From Eq.~\eqref{eq:rE1}, however, we can see that half of the electric
charge is given by the baryon charge, so the influence of the latter
is 50\%.
Nevertheless, we can see from Fig.~\ref{fig:rpE} that the electric
charge radius has quite a different behavior in the parameter space
than the baryon charge radius.

We can see that the electric charge radius is generally quite a bit
larger than the baryon charge radius. This is physically reasonable. 
In general, it is seen that the BPS-Skyrme term helps decreasing the 
electric charge radius, but not nearly enough to reach the
experimentally measured value.
We can also see that a larger value of the pion mass again helps
decreasing the radius.

\begin{figure}[!htp]
  \begin{center}
    \mbox{
      \subfloat[$(m_2,c_6)=(0,0.5)$]{\includegraphics[width=0.49\linewidth]{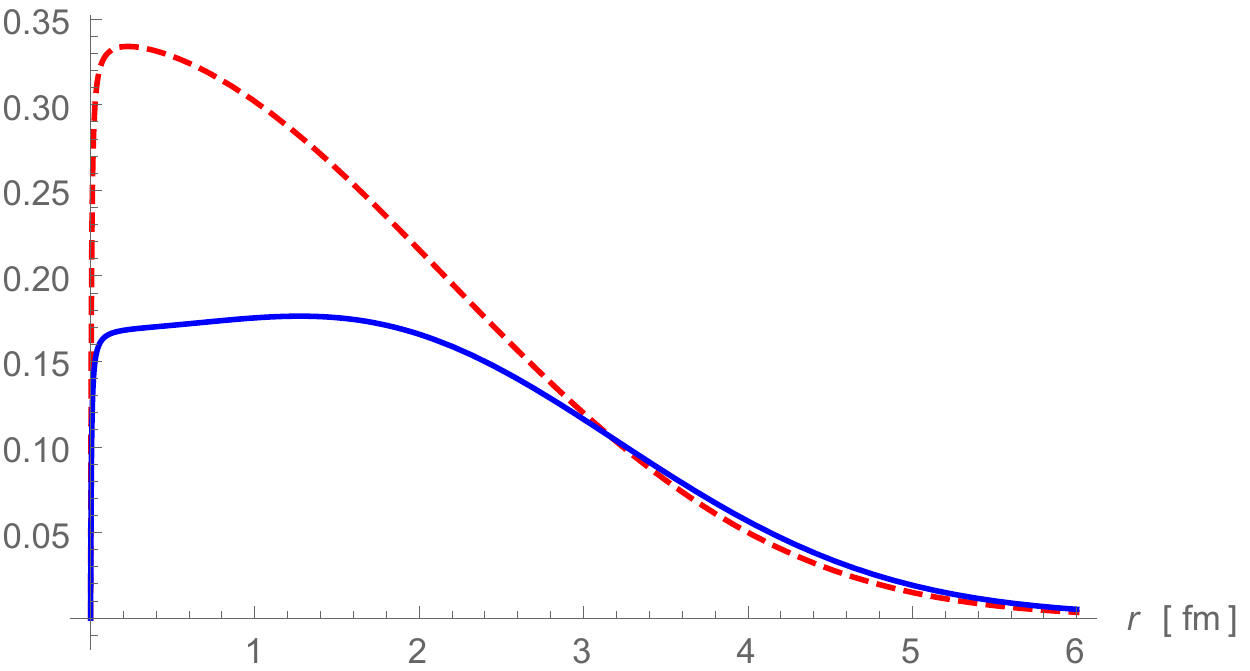}}
      \subfloat[$(m_2,c_6)=(0.8,0.5)$]{\includegraphics[width=0.49\linewidth]{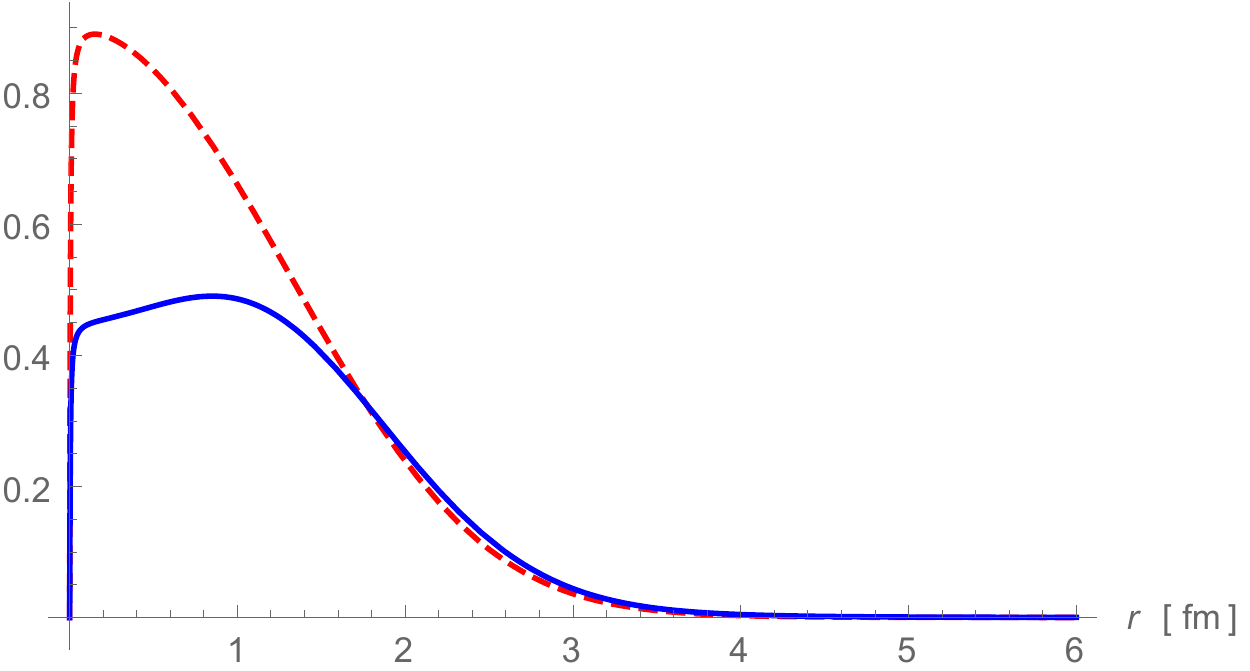}}}
    \mbox{
      \subfloat[$(m_2,c_6)=(0,0)$]{\includegraphics[width=0.49\linewidth]{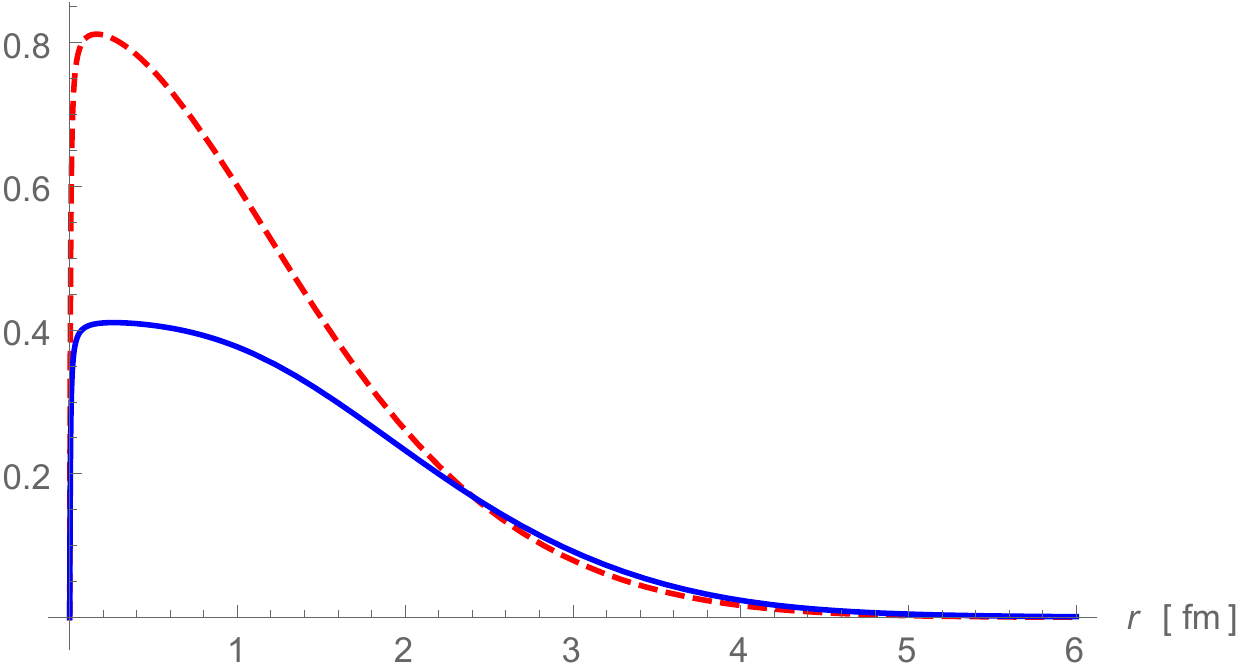}}
      \subfloat[$(m_2,c_6)=(0.8,0)$]{\includegraphics[width=0.49\linewidth]{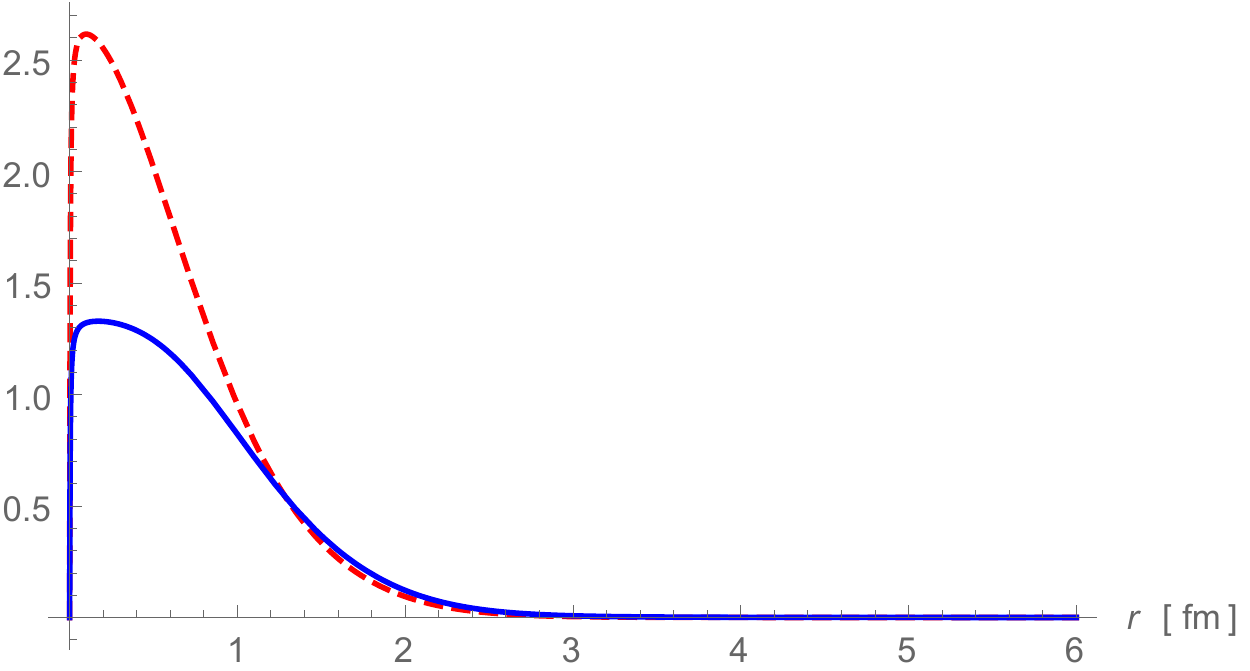}}}
  \end{center}
  \caption{Baryon and electric charge densities,
    $\frac{1}{2\pi^2}\mathcal{B}^0$ as red dashed lines and $\rho_E$ as
    blue solid lines, in various points of the parameter space:
    $(m_2,c_6)=(0.5,0),(0.5,0.8),(0,0),(0,0.8)$. 
  }
  \label{fig:rps}
\end{figure}

As an example, we illustrate the baryon and electric charge densities
in four points of the parameter space in Fig.~\ref{fig:rps}.

\subsection{Axial coupling}

\begin{figure}[!htp]
  \begin{center}
    \mbox{
      \subfloat[]{\includegraphics[width=0.49\linewidth]{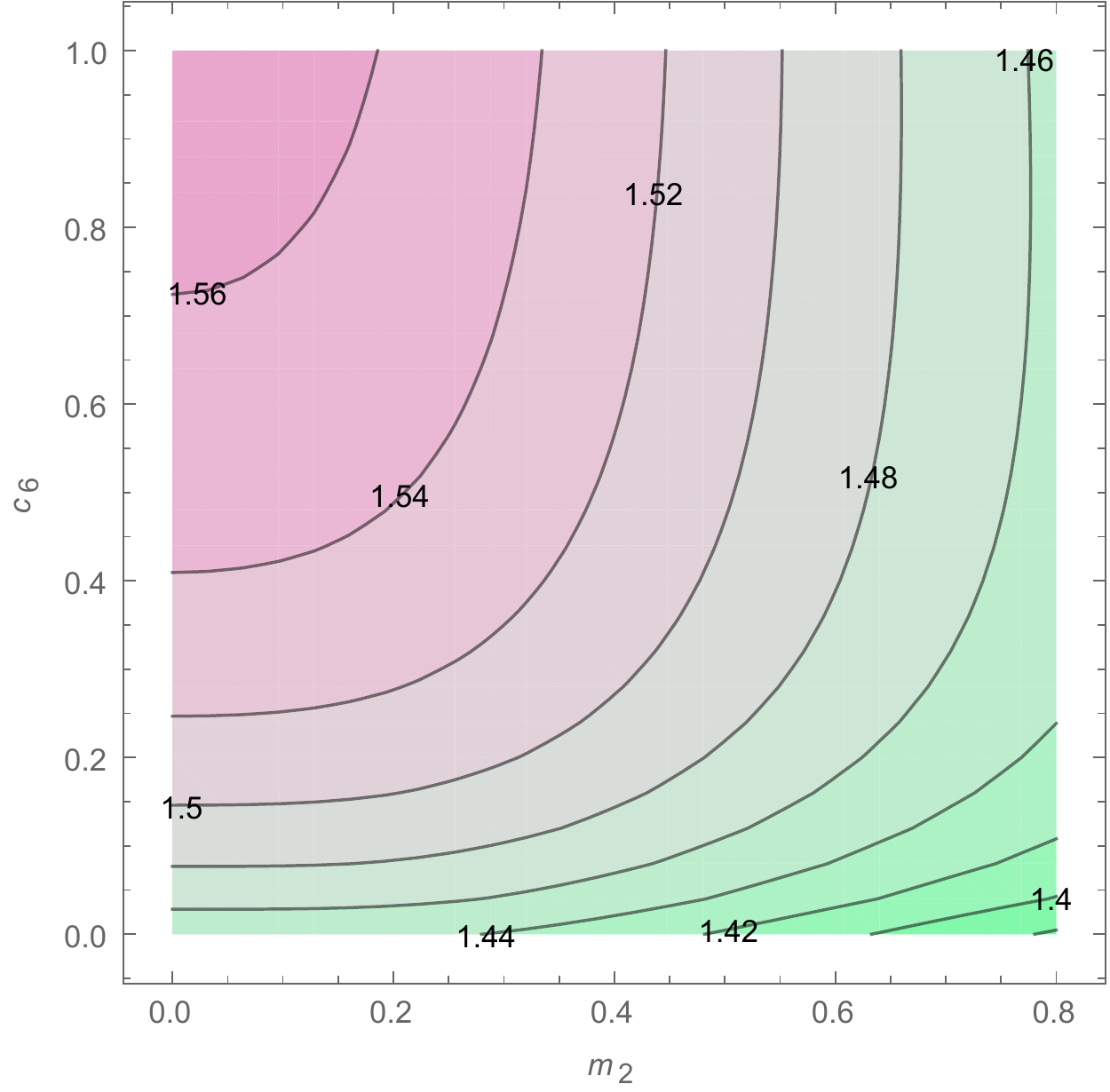}}
      \subfloat[]{\includegraphics[width=0.49\linewidth]{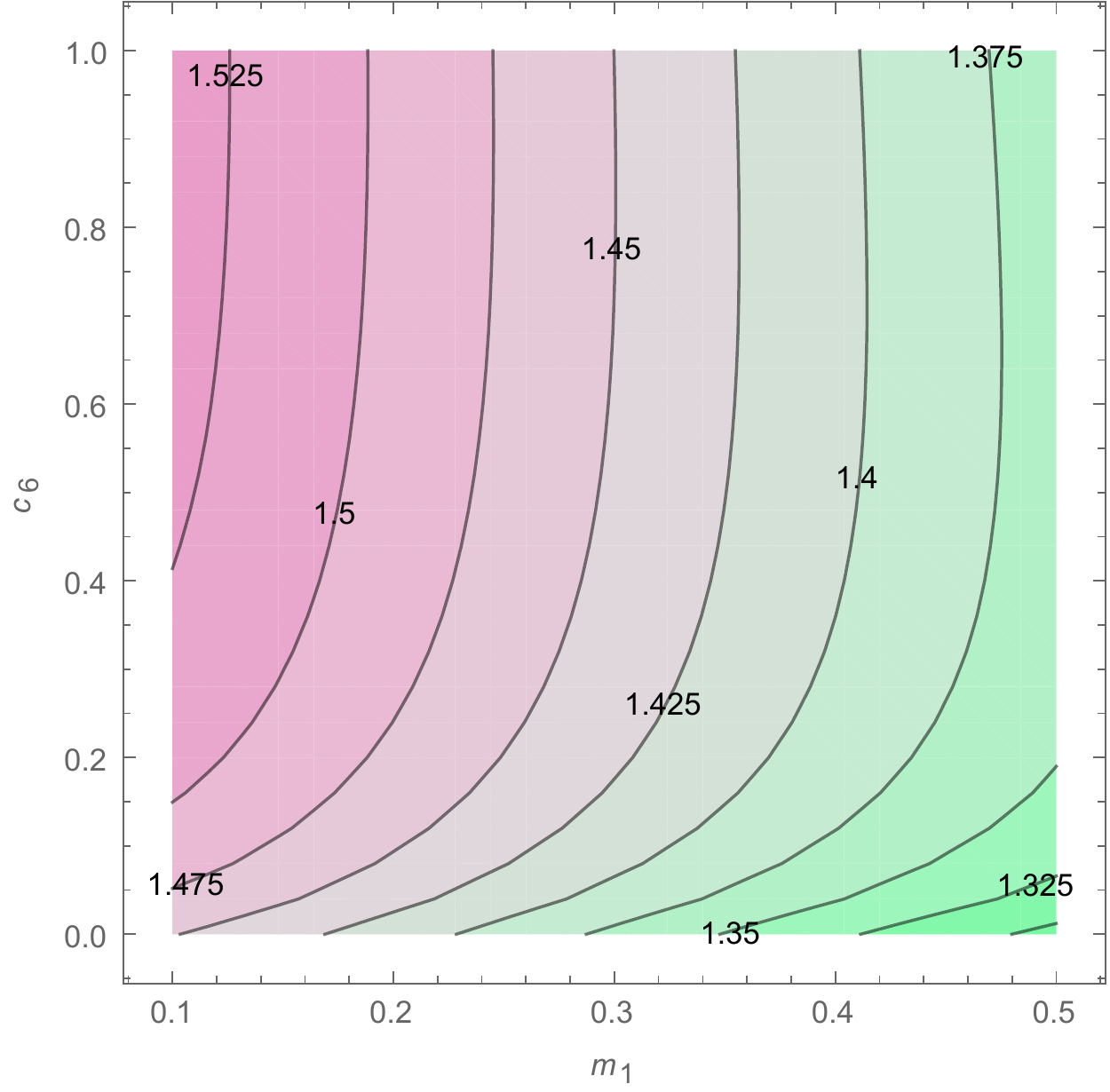}}}
  \end{center}
  \caption{Axial coupling, $g_A$, in (a) the $(m_2,c_6)$ plane for
    $m_1=0.25$ and (b) the $(m_1,c_6)$ plane for $m_2=0.7$. 
  }
  \label{fig:gA}
\end{figure}

The final observable that we will present in this paper is the axial
coupling, following Ref.~\cite{Adkins:1983ya}, but again with the
inclusion of the contribution from the BPS-Skyrme term, see
Eq.~\eqref{eq:gAtilde}.
The result of the axial coupling in the chosen parameter space is
shown in Fig.~\ref{fig:gA}.

Although the original calculation in Ref.~\cite{Adkins:1983ya} yielded
a value of $g_A$ almost half the size of the experimentally measured
value, our model overestimates this observable.
This can be traced back to the value of the Skyrme-term coefficient,
$e$, which is much smaller in our model when calibrated against the
mass and size of Helium-4.
Although we overestimate the value of the axial coupling, we can see
that if we were to increase $m_2$ beyond the chosen parameter space
for large values of $c_6$ and possibly also increase the pion mass,
the value could conceivably decrease to the right orders of magnitude
of the experimentally measured value of about 1.27 \cite{PDG2015}.
That part of parameter space is, however, inaccessible in this paper
because we cannot know if the cubic symmetry of the 4-Skyrmion is
retained there or not.
That requires the full PDE calculations and will be interesting land
to cover in the near future.

\subsection{Summary of the results}

The loosely bound potential decreases the classical and total binding
energies \cite{Gudnason:2016mms,Gudnason:2016cdo} and in this paper we
have shown that the BPS-Skyrme term with an order-one coefficient can
help decreasing it further.
In particular, for $m_2=0.8$: increasing $c_6$ to $c_6=1$, decreases
the binding energy by 0.9\% (after taking the calibration into
account) and it worsens the values of the nucleon mass, axial
charge and spin contribution to the nucleon; it improves the value of 
the proton charge radius; 
has no effect on the $\Delta$ mass and pion decay constant; and it
overshoots the value of the pion mass.
The tendencies seen in our results in this section, indicate that
enlarging the parameter space using full PDE calculations, one may
find solutions with lower and more realistic binding energies than
found in this paper.

\section{Discussion}\label{sec:discussion}

In this paper we have studied the most accessible part of the
parameter space of the loosely bound Skyrme model with the BPS-Skyrme
term and the second-order potential providing the lowest binding
energies.
For the first time, we have used the rational map approximation for
the Skyrme model quite far outside of its common range; that is, we
have turned on both the loosely bound potential and the BPS-Skyrme
term.\footnote{The rational map is often used in the (pure) BPS-Skyrme
  model \cite{Adam:2010fg,Adam:2010ds} where it is not an
  approximation due to volume-preserving diffeomorphism invariance. } 
Interestingly, the same function of the rational map applies to the
BPS-Skyrme term as appeared already in the normal Skyrme term. 
For the loosely bound potential we could compare our results to those
of Ref.~\cite{Gudnason:2016cdo} and calculate the systematic error by
using the rational map approximation: about 1.5-2.5\%, with an
approximate linear increase as function of the coefficient of the
loosely bound potential, $m_2$. 
We have found the classical binding energy as low as 1.8\% and total
binding energy down to about 5.3\%, in the covered part of parameter
space.
With the results of Ref.~\cite{Gudnason:2016cdo} in mind, we expect
that in the part of parameter space where the pion mass is large, the
coefficient of the loosely bound potential is allowed to be larger
than covered here and then turning on a sizable BPS-Skyrme term will
result in even lower binding energies.

Another result that we have calculated in this paper is the
contribution from the BPS-Skyrme term to the axial coupling. It
generally increases the value of the coupling, which is a wanted
feature in light of the original results \cite{Adkins:1983ya};
however, with our calibration the coupling turns out to be slightly
too large compared with its experimental value.
Fortunately, the tendency of the results in this paper points in the
direction that the region of parameter space where we may lower the
binding energy further than what we have covered in this paper, may
lead to a lower and perhaps quite reasonable value of the axial
coupling.

If we pick the vanilla point in our covered parameter space -- based
entirely on the binding energy -- we get a total binding energy of
5.3\%, the pion decay constant at 71 MeV, $e\sim 4.5$, the nucleon
mass at 960 MeV, the mass of the Delta resonance at 1095 MeV, the pion
mass at 75 MeV, the proton charge radius at 1.25 fm, and finally, the
axial coupling at 1.53.
A comparably good point in parameter space gives us instead, a total
binding energy of 5.5\%, the pion decay constant at 65 MeV,
$e\sim 4.6$, the nucleon mass at 960 MeV, the mass of the Delta at
1100 MeV, the pion mass at 150 MeV, the proton charge radius at 1.15
fm and finally, the axial coupling at 1.46.

The first thing that can be done from now is to explore the parameter
space of the loosely bound Skyrme model with the BPS-Skyrme term
turned on using full PDE calculations.
This will first of all confirm our results here (we expect 1-2\% error
and perhaps less after correcting for the systematic error at the
$c_6=0$ slice of parameter space), but it will also allow one to go 
farther into unexplored regions of parameter space, which are expected
to contain solutions with even lower binding energies than found in
this paper.
There are two directions to explore. The first is towards the near-BPS
regime of the BPS-Skyrme model which in our language means very large
values of $c_6$ and $m_2$. The problem with this approach is the
technically difficult numerical calculations that need to be
tackled. 
The other direction to search in, is to take a large pion mass; crank
up $m_2$ to the boundary of where the cubic symmetry of the 4-Skyrmion
is lost and then increase $c_6$ (the BPS-Skyrme term) and again
increase $m_2$ as much as possible. Continue this loop until the
binding energy is of the right order of magnitude (or until a
technical problem occurs).

Another interesting future direction would be to consider the
derivation of the BPS-Skyrme term in the framework of partially
bosonized QCD at large $N_c$, along the lines of
Ref.~\cite{Zaks:1985cv}. 

Let us mention an obstacle that we have not mentioned so far.
We are pursuing the lowest possible binding energies in a
generalization of the Skyrme model in order to match experimental
data. However, even if we can reduce the classical binding energy
arbitrarily, say down to $\varepsilon$, then the tendency is that the
method of adding the quantized spin-isospin contribution to the
classical mass by itself yields about 2-3\% binding energy.
Thus to reach agreement with experiments, we need a classical binding
energy of about $-$(1-2)\%, which means that the classical solutions
are unbound and thus impossible.
It is plausible that vibrational modes could play a role in this
problem (like for ${}^7$Li and ${}^{16}$O in
Refs.~\cite{Halcrow:2015rvz,Halcrow:2016spb,Davies:2009zza}), so that
the 4-Skyrmion would receive a nonrotational quantum contribution and
thus lower its binding energy. 
This issue of the large size of the spin-isospin contribution to the
1-Skyrmion has been addressed recently in Ref.~\cite{Adam:2016drk},
where it was suggested by an indirect argument of relating the spin
contribution to the nucleon mass and the moment of inertia of the
nucleon, that the spin contribution should actually be more than twice
as large as the standard Skyrme model (direct) argument suggests.
The spin contribution to the mass of the nucleon needed should, by
comparing the nucleon to Helium-4, be as low as 7.2 MeV.
However, by the indirect argument of Ref.~\cite{Adam:2016drk} it could
be as large as 16 MeV, reducing a lot of tension in the Skyrme model.
For consistency, however, this requires \emph{some} quantum
contribution to the 4-Skyrmion.

\subsection*{Acknowledgments}

S.~B.~G.~thanks the Recruitment Program of High-end Foreign
Experts for support.
This work was supported by the National Natural Science Foundation of
China (Grant No.~11675223).

\end{document}